\documentclass[showpacs,printnumbers,amsmath,amssymb, aps, pra,twocolumn]{revtex4-1}

\usepackage{graphicx}
\usepackage[normalem]{ulem}
\usepackage{dcolumn}
\usepackage{bm}
\usepackage{xspace} 
\usepackage{color}
\usepackage{enumerate}
\usepackage{subfigure}

\newcommand{\ket}[1]{|{#1}\rangle}
\newcommand{\ave}[1]{\langle{#1}\rangle}               

\def\Rb87{$^{87}\text{Rb}$}
\def\0{\ket{0}}
\def\1{\ket{1}}

\usepackage{hyperref}
\hypersetup{
citecolor=blue,
 colorlinks=true,
 urlcolor=magenta,
}

\begin{document}


\title{Supersensitive quantum sensor based on criticality in an antiferromagnetic spinor condensate}

\author{Safoura S. Mirkhalaf}
\affiliation{Institute of Physics PAS, Aleja Lotnikow 32/46, 02-668 Warszawa, Poland}
\author{ Luca Lepori} 
\affiliation{Istituto Italiano di Tecnologia, Graphene Labs, Via Morego 30, I-16163 Genova, Italy}
\author{Emilia Witkowska}
\affiliation{Institute of Physics PAS, Aleja Lotnikow 32/46, 02-668 Warszawa, Poland}
\email{mirkhalafsa@gmail.com}

\date{\today}

\begin{abstract}
We consider an antiferromagnetic Bose-Einstein condensate in a traverse magnetic field with a fixed macroscopic magnetization. The system exhibits two different critical behaviors corresponding to transitions from polar to broken-axisymmetry and from antiferromagnetic to broken-axisymmetry phases depending on the value of magnetization. We exploit both types of system criticality as a resource in the precise estimation of control parameter value. We quantify the achievable precision by the quantum Fisher information. 
We demonstrate supersensitivity and show that the precision scales 
with the number of atoms up to $N^4$ around critically. In addition, we study the precision based on the error-propagation formula providing the simple-to-measure signal which coincide its scaling with the quantum Fisher information. Finally, we take into account the effect of non-zero temperature and show that the sub-shot noise sensitivity in the estimation of the control parameter is achievable in the low-temperature limit.
\end{abstract}

\maketitle
\section{Introduction}
%
Properties of a system can change dramatically trough a small change in a control parameter during a phase transition. Phase transitions can be of classical or quantum nature. An example of the classical transition is the ice-water-vapor transition for water in the H$_2$O system or the ferromagnetic-paramagnetic transition in solid-state materials, with temperature as the external parameter in both cases. On the other hand, the quantum phase transitions occur at zero temperature and are induced by a change of a Hamiltonian parameter. Phase transitions are classified accordingly to the basic Ehrenfest classification~\cite{Jaeger} as first- and second-order. However, other classifications are also widespread~\cite{Blundell}. A first-order phase transition is characterized by the coexistence of two stable phases when the control parameter is within a certain range. On the other hand, a second-order phase transition is characterized  by a massless spectrum, inducing power law scaling for correlations and the notion of universality, that in turn results in a bunch of critical exponents quantifying how fast the system changes around criticality.  

At the heart of quantum metrology lies the idea of  parameter estimation using a quantum resource. The best precision in the estimation of a particular parameter is quantified by the quantum Fisher information (QFI), related to the distinguishability of a quantum state from a neighbor state in a geometrical space~\cite{caves1994}. It has already been  recognized \cite{zanardi2008} that criticality is considered as a perfect resource for parameter estimation. This happens because quantum states around criticality differ strongly from each others, although the control parameter driving the transition varies by a small amount.

To date, the role of quantum criticality for  parameter estimation has been investigated in the Lipkin-Meshkov-Glick~\cite{Paris2014}, Dicke \cite{Paris2016,Paris2019}, 
bosonic Josephson junction \cite{Pezze2019} and many other quantum models \cite{Rams2018}. 
The first experiment demonstrating the high sensitivity in parameter-estimation around citicality was reported recently for the system composed of condensed atoms in a double-well potential~ \cite{Pezze2019}.
While the majority of these works are devoted to examining the criticality around the second-order phase transitions, only a few works concern first-order ones \cite{Sanpera2014,Sanpera2018,Vicari2018}. However, due to the much more drastic change of the ground state properties around a first-order transition, it can be interesting to investigate its relevance for control parameter estimation hiring the states as a quantum resource. Therefore, here we consider a system of spin-1 Bose-Einstein condensate which is presenting both a first- and a second-order phase transition, depending on the parameters of the system.

\begin{figure}[]
	\centering {\includegraphics[width=\columnwidth]{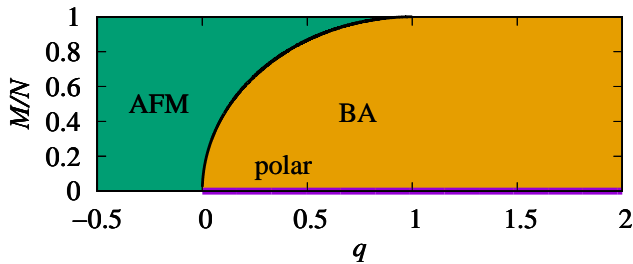}} 
	\caption{(Color online) Mean-field phase diagram of the antiferromagnetic spinor condensate under the single mode approximation and for fixed magnetization $M$ hosts three different phases~\cite{dressing2015}. \emph{The antiferromagnetic (AMF) phase:} the components $m_F=\pm 1$ coexist . \emph{The broken-axisymmetry (BA) phase:} atoms occupy all three Zeeman components. \emph{The polar phase:} all atoms are in the $m_F=0$ Zeeman component. In general, the ground state is a superposition of Fock states. However, when $q\ll q_c$, the ground state of AFM phase is $|(N+M)/2,0,(N-M)/2\rangle$. While for $q\gg q_c$, the ground state of BA phase reads $|M,N-M, 0\rangle $. Particularly in the polar phase the ground state is $|0,N,0\rangle$. The solid black line shows the position of critical point $q_c=1-\sqrt{1-(M/N)^2}$. The quantum phase transition is the first-order from the polar to the antiferromagnetic phases, it occurs when the magnetization tends to zero. In other cases, the transition from the broken-axisymmetry to the antiferromagnetic phases is of the second-order. }
	\label{fig:fig1}
\end{figure}

Spinor Bose-Einstein condensates (BECs) are composed of $N$ atoms in several Zeeman energy levels with a given hyperfine spin $F$ numerated by the magnetic number $m_f\in [-F,F]$. Here we concentrate on $F=1$. The system possess an additional spin degree of freedom which leads to a range of phenomena absent in a scalar BEC. The longitudinal magnetization $M$, which is a difference in the occupations of the $m_f=1$ and $m_f=-1$ components, is approximately conserved in the system and acts as an independent external parameter. This conservation law comes from the spin rotational symmetry of contact interactions when dipole-dipole interactions are neglected. The global ground state of the $F=1$ system is classified on ferro or antiferromagnetic, depending on the sign of spin-dependent interactions. The structure of the ground state of a homogeneous system results from the competition between spin-dependent interactions (dominant at low magnetic fields) and the quadratic Zeeman energy (dominant at large magnetic fields) which gives rise to emergence of two different phases and to a critical point which lies in between them. The position of the critical point depends on the value of magnetization as depicted in Fig.~\ref{fig:fig1}. More importantly for the purpose of our work, the order of the phase transition does depend on the magnetization value as well. It is of the first-order when the magnetization tends to zero \cite{Raman2011,Raman2017}, and second-order for the macroscopic one \cite{dressing2014,dressing2015,Gerbier2012}. 

The purpose of the paper is to perform a comprehensive study of the metrological usefulness of the two types of criticality appearing in the antiferromagnetic condensate. We concentrate on the finite size system (a few thousands of atoms) in which the spatial and internal degrees of freedom can be decoupled. We quantify the metrological usefulness by the quantum Fisher information determined by the fidelity susceptibility \cite{zanardi2008}. Our numerical method is based on the exact diagonalization of the system Hamiltonian and on the consequent evaluation of the QFI for the ground state. We relate the scaling of the QFI with the critical exponents {\cite{Rams2018}} for macroscopic magnetizations, showing that it scales with the system size as $~N^{4/3}$. Our results confirms the general treatment provided in \cite{Rams2018}. In the case of zero magnetization, when the phase transition is the first-order, we found its scaling with the system size to be $N^4$. We confirmed the numerical results by the analytical perturbative approach. We show that the QFI around the first-order phase transition is much more prominent, at least in the zero temperature case. In addition, we show that the precision in the estimation of the control parameter can be achieved by using a simple signal, which is introduced as the atomic population in the $m_f=0$ Zeeman component. Specifically, we evaluate the  estimation precision employing the error propagation formula and confirm that the scaling of its inverse with $N$ coincides with the scaling of the QFI for any magnetization as expected~\cite{Pezze2019}. 
The extensive use of the error propagation formula, simpler to obtain than QFI also in experiments, to estimate the sensitivity both at a first- and at a second-order phase transition is one of the central points of the present work. Finally we consider the effect of a non-zero temperature showing that the value of the QFI drops down faster for the zero magnetization than the macroscopic one. However, the sub-shot noise scaling of the QFI can be still possible when the temperature is lower or of the order of the energy gap.

The paper is organized as follows. In Section \ref{sec1}, we introduce the model and review the characteristic properties of its phase diagram. In Section \ref{sec2}, we provide the basics of the estimation theory around criticality. Next, we present our results in detail for zero and non-zero temperature in Sections \ref{sec3} and \ref{sec4}, respectively. The concluding remarks and summary are given in Section \ref{conclusion}.

\section{The spin-1 system}\label{sec1}

We consider the spin-1 ($F=1$) atomic Bose-Einstein condensate in the presence of a homogeneous transverse magnetic field $B$. The system is conveniently described by the vector $\vec{\hat{\Psi}}=(\hat{\Psi}_1,\hat{\Psi}_0,\hat{\Psi}_{-1})^T$  which components correspond to the atoms in the corresponding Zeeman states numerated by the quantum magnetic number $m_f=0,\pm 1$. 
We consider the finite-size system composed of a few thousands of atoms in which generation of spin domains are energetically costly. It is convenient to work under the \emph{single-mode approximation} (SMA) in which all atoms in the three Zeeman modes share the same spatial wave function $\phi(\boldsymbol{r})$ \cite{Ueda2012}. Then, the external and internal spin degrees of freedom can be decoupled and the components of the vector are defined as $\hat{\Psi}_{mf}=\phi(\boldsymbol{r})\hat{a}_{mf}$, where $\hat{a}_{mf}$ is the bosonic annihilation operator of an atom in the $m_f$-th Zeeman state. Consequently, the Hamiltonian casts in the following form \cite{Rotor2010,Duan2013} 
\begin{eqnarray}\label{H}
\frac{\hat{H}(q)}{c}=\frac{1}{2N}\hat{J}^2-q\hat{N}_0,
\end{eqnarray}
and consists of two terms: the first one refers the non-linear contact interactions between pairs of atoms while the second term shows the effect of quadratic Zeeman shift on the energy levels. In Eq.(\ref{H}), $\hat{J}^2$ is the total spin operator which can be defined in terms of the spin-1 matrices (see appendix \ref{app1}), $\hat{N}_{mf}$ is the occupation number operator of atoms in the $m_f$ Zeeman state, the total atom number $N$ is the eigenvalue of $\hat{N}=\sum_{m_f}\hat{N}_{m_f}$ and ${c}=Nc_2\int d{\boldsymbol{r}}|\phi(\boldsymbol{r})|^4$ with the spin dependent interactions coefficient $c_2$ defined in terms of s-wave scattering lengths~ 
\footnote{The explicit form of $c_2$ is given as $c_2=4\pi\hbar^2(a_0-a_2)/3m$, where $m$ is mass of each particle and $a_0$ ($a_2$) are the s-wave scattering lengths for spin-1 atoms colliding in symmetric channels of total spin $J=0$ ($J=2$).}. The positive (negative) sign of $c$ represents the antiferromagnetic (ferromagnetic) nature of interactions \cite{Ueda2012}. 
The coupling constant $q$ gives the strength of the quadratic Zeeman energy. In fact, the parameter $q$ can be a sum of two terms, $q=q_{\rm B} + q_{\rm M}$, as it can be changed using the magnetic and off-resonant microwave dressing fields denoted by $q_B$ and $q_M$, respectively \cite{dressing2014,dressing2015}. The value of $q$ can be therefore tuned between negative and positive values.

The Hamiltonian (\ref{H}) preserves the $z$ component of the total spin operator, $\hat{J}_z=\hat{N}_{+1}-\hat{N}_{-1}$ due to  $[\hat{H},\hat{J}_z]=0$. Therefore, the eigenvalues of $\hat{J}_z$ which are $M=-N,-N+1,...,N$, being the magnetization, can be used to label the Hamiltonian eigenbasis (more details in Appendix \ref{app1}). This is justified based on the fact that the spin-dependent interaction has rotational symmetry as long as the spin-1 system is isolated from its environment and dipolar interactions are neglected \cite{Gerbier2012}. This is also the main reason why the linear Zeeman energy plays no role in the Hamiltonian (\ref{H}), and it acts only as a constant shift on the energy levels. 

Our numerical method is based on the exact diagonalization of the Hamiltonian (\ref{H}) and is described in details in the Appendix \ref{app2}. For convenience, we consider even values of $N$, non-negative values of $M$, and the antiferromagnetic interactions ($c>0$) which can be realized with a condensate of Sodium-23 atoms in the $F=1$ or $F=2$ manifolds. For the case of trapping atoms in a flat 
trap of volume $V$, we can assume a homogeneous density for the condensate, such that $\phi(\boldsymbol{r})=\frac{1}{\sqrt{V}}$. Consequently, $c=c_2\frac{N}{V}$. Here, $\frac{N}{V}$ is the density of system which is maintained as a fixed parameter. In the following, we use $c$ as the energy unit.

\begin{figure}[]
	\centering {\includegraphics[width=1\columnwidth]{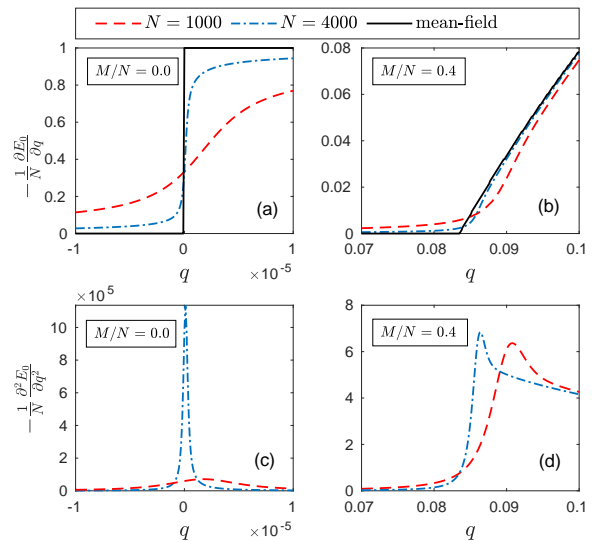}} 
	\caption{(Color online) Upper panel: the first derivative of the ground state energy of the Hamiltonian (\ref{H}) versus $q$ obtained numerically using exact diagonalization method for $N=1000$ (dash-dotted blue) and $N=4000$ (dashed red), compared to the mean-field results (the black solid line). Lower panel: the second derivative of the ground state energy versus $q$ for the same parameters. Left and right columns correspond to $M/N=0$ and $M/N=0.4$, respectively.}
	\label{fig:fig2}
\end{figure}  

It has been already discussed in the literature that the Hamiltonian (\ref{H}) exhibits both the first- and second-order phase transitions at critical values of the external parameter $q$~\cite{Ueda2012,Ueda2013}. In particular for the case of zero magnetization, the transition occurs between the longitudinal polar and broken-axisymmetry phases while in the case of macroscopic magnetization the transition is between antiferromagnetic and broken-axisymmetry ones. We provide the characteristics of particular phases and expressions for the corresponding ground states in caption of Fig.~\ref{fig:fig1}. Moreover, using the fractional occupation number in the $m_f=0$ state $n_0=\langle \hat{N}_0\rangle /N$ as the order parameter, one can define the critical value of $q$ as $q_c=1- \sqrt{1-(M/N)^2}$ for a given $M/N$ in the thermodynamic limit using mean-field approach~\cite{Zhang2003}. The corresponding phase diagram of the antiferromagnetic condensate has been explored experimentally~\cite{dressing2014,dressing2015,Gerbier2012} and the agreement with theoretical predictions has been noticed. 

In the many body systems in thermodynamic limit, an abrupt continuous (discontinuous) change of the first derivative of the ground state energy (at zero temperature) around criticality is observed. This behaviors mark the continuous second (discontinuous first) order phase transition. The radical change of the derivative of the ground state energy $E_0$ is also linked to the abrupt changes of the order parameter based on the Hellmann-Feynman theorem which gives $n_0 \equiv-\ave{\frac{1}{N}\frac{\partial \hat{H}(q)}{\partial q}}=-\frac{1}{N}\frac{\partial E_0(q)}{\partial q}$~\cite{Ming2018} by considering our Hamiltonian (\ref{H}). In Fig.~\ref{fig:fig2} we show variations of the first (upper) and of the second (lower) derivative of the ground state energy of the Hamiltonian (\ref{H}) for the finite size system with $N=1000,\, 4000$. 
In the case of macroscopic magnetization (right column) the first derivative of the ground state
changes continuously while the second derivative of energy exhibits an abrupt but continuous change around criticality. The second derivative shows a discontinuous behavior in the thermodynamic limit. On the other hand for zero magnetization, the first derivative of $E_0$ shows a continuous but sudden variation (left column). 
This variation trends to a discontinuity when approaching the thermodynamic limit. The peak of the second derivative of energy is much sharpened around criticality,  It is worth to notice, that for $M=0$ and $q\sim 0$ the ground state gives $\ave{\hat{{N}}_0}=N/3$ \cite{Bigelow1998}. In this case, all the higher-order derivatives of the ground state energies are discontinuous. In general, we conclude that the quantum phase transitions are quite smooth, due to the finite sizes of the considered system. {While the mean-field works in the thermodynamic limit, for typical ultracold gas experiment with average size ensembles, the mean-field results does not hold necessarily}. In this case, the variation of population observables (such as $n_0$'s) should be extracted in the full quantum approach \cite{Gerbier2012}.
For the finite-size condensate of $N$ spin-1 atoms, the value of $q_c$ depends on the ratio $M/N$ and $c$, at least to some extents. 
In order to drive the system throughout the critical region one can tune the control parameter $q$ by external magnetic field or microwave dressing from negative to positive values. This can be used to estimate the value of $q$. 

Indeed, it has already been discussed that the family of quantum states around a critical point can be used as a resource for quantum sensing \cite{zanardi2008}. This is possible because a small variation of control parameter around criticality leads to a remarkable change in the properties of these states. In the following, we analyze both types of criticality in the antiferromagnetic spin-1 system, 
and show that the precision in the estimation of the coupling constant $q$ is greatly enhanced as compared to the noncritical regions.
To this end, in the next section we briefly present the relation between criticality and metrology. In this spirit, we describe the quantum Fisher information as the essential parameter which provides a bridge between these territories. 

\section{Quantum estimation theory around criticality}\label{sec2}

A quantum phase transition concerns a radical change in the ground states of a particular Hamiltonian at a specific critical point. It has been proved \cite{zanardi2008} that while varying a control parameter drives the system into different phases, 
this can be used to enhance the precision in the estimation of the control parameter by its own.
This means that criticality can be a resource in quantum metrology. Here, we recall the main ingredients of the critical metrology formalism which is based on the definition of the quantum Fisher information.

Let us consider the generic Hamiltonian form
\begin{eqnarray}\label{H2}
\hat{H}(q)=\hat{H}_0 + q \hat{H}_q ,
\end{eqnarray}
which ground state $|\psi_0(q)\rangle$ exhibits a quantum phase transition at a critical value of $q=q_c$ (at zero temperature). This means that, the ground state of the above Hamiltonian have a drastic change varying from $|\psi_0(q)\rangle$ to $|\psi_0(q + \delta_q)\rangle$ for a small variation of $\delta_q$ around the quantum critical point (QCP) at $q_c$. The physical quantity that is used to evaluate the difference between the two pure quantum states, including ground state, is defined by the fidelity~\cite{Jozsa1994}
\begin{eqnarray}\label{fid}
\mathcal{F}=\left| \langle\psi_0(q)|\psi_0(q+\delta_q)\rangle \right|.
\end{eqnarray} 
The relation between the fidelity and the quantum Fisher information $F_q$ is \cite{caves1994,Cozzini2007,You2007,You2015} 
\begin{eqnarray}\label{QFI_fid}
F_q=-4 
\frac{\partial^2 \mathcal{F}}{\partial \delta_q^2} |_{\delta_q=0}.
\label{QFI_def}
\end{eqnarray}
\noindent More explicitly \cite{zanardi2008}
\begin{eqnarray}
F_q(|\psi_0(q)\rangle)
&=&4\left[ \langle\partial_q\psi_0|\partial_q\psi_0\rangle-|\langle\partial_q\psi_0|\psi_0\rangle|^2 \right] \label{QFI1} \\
&=&4\sum_{n\neq 0}\frac{|\langle\psi_0|\hat{H}_q|\psi_n\rangle|^2}{(E_0-E_n)^2}. \label{QFI}
\end{eqnarray}
Here, $|\psi_n\rangle$ and $E_{n}$ refers to the $n$-th excited eigenstates and eigenvalues of (\ref{H2}), respectively, and $\partial_q\equiv \partial/\partial q$. Moreover, the equation (\ref{QFI1}) can be obtained by Taylor expanding the state  $|\psi_0(q+\delta_q)\rangle$ around $|\psi_0\rangle$ up to the second order in $\delta_q$ and excluding the first derivate term due to the normalization condition, $\partial\langle\psi_0|\psi_0\rangle/\partial q=0$~\cite{Zanardi2007}.

It is important to note, that the formula (\ref{QFI1}) is valid if the first derivative of the ground state exists. In the case of the first-order quantum phase transition, due to the level crossing the first derivative of a ground state is discontinuous. However, in this work, we focus on a finite size system and therefore the level crossing changes to avoided level crossing. As a result the definition (\ref{QFI1}) is still valid for the case of first-order phase transition~\cite{Chen2008}. In addition, as discussed in~\cite{Paris2019,Dominik2017}, the QFI and the Bures metric correspondence is not broken provided that the rank of the GS density matrix is not changed at the critical region \footnote{The rank of a matrix is the maximum number of linearly independent row vectors (or equivalently column vectors) of a matrix \cite{Vaughn2007}}. This also works for our case that the quantum state of the system remains pure with rank 1 for zero temperature and of rank 2 in the finite temperature case \cite{Paris2019,Dominik2017}.

The QFI is related to the geometrical distinguishability of quantum states separated by a small variation of $q$. Consequently, its value is significantly increased around criticality. This is easily observed from the QFI (\ref{QFI}) since one of the excited eigenvalues approaches the ground state at the QCP~\cite{zanardi2008}. On the other hand, the QFI is connected to the precision in the estimation of the $q$ parameter. In (\ref{H2}), one may consider the unknown coupling constant as an imprinted phase to be measured~\cite{Rams2018}. Since there is no direct observable corresponding to coupling constants, we cannot measure its value by just using the conventional approach in quantum mechanics, that is evaluating the expectation value of an observable on a particular state. Therefore, the problem of measuring $q$ turns into an estimation problem \cite{Paris2009}. The ultimate bound of estimation, called the Quantum Cramer-Rao bound (QCRB), is set by the inverse of the QFI
\begin{eqnarray}\label{QCRB}
\delta q^2 \geq \delta q^2_{QCRB} = \frac{1}{F_q}
\end{eqnarray}
Therefore, the precision in the estimation of $q$ is significantly improved at criticality, implying that it is a resource in estimation theory~\cite{zanardi2008,Zanardi2007}.

In the case of mixed states $\hat{\rho}(q)=\sum_n w_n |\psi_n\rangle\langle\psi_n|$, the fidelity  (\ref{fid}) is replaced by the more general definition \cite{Uhlmann1976,Jozsa2007}
\begin{eqnarray}\label{fidel}
\mathcal{F}={\rm tr}\left[\sqrt{\hat{\rho}(q)}\hat{\rho}(q+\delta_q)\sqrt{\hat{\rho}(q)}\right]^{1/2}
\end{eqnarray}
which can still be exploited, by using (\ref{QFI_fid}) to obtain the QFI. The explicit result for a single parameter estimation is 
\cite{Paris2009,Paris2019B}
\begin{eqnarray}\label{fidel-p}
F_q=2\sum_{n,m}\frac{|\langle\psi_n|\partial_q\hat{\rho}_q|\psi_m\rangle|^2}{w_n+w_m}.
\end{eqnarray}
for finite number of particles and continuous phase transition \cite{Paris2019B}. 

As mentioned before, the QFI gives the upper bound of sensitivity. However, it is not always easy to find the optimal measurement to saturate the QCRB. Moreover, it is not straightforward to extract the QFI experimentally. This refers to the fact that in practice in order to find the QFI, one needs the full tomography of $\hat{\rho}(q)$ and $\hat{\rho}(q+\delta_q)$. This process is not easy to be implemented for the large systems. Therefore, it is convenient to consider the precision given by the error-propagation formula defined as the signal-to-noise ratio \cite{Rams2018}
\begin{eqnarray}
\delta q^2=
\frac{\Delta^2{\hat{\mathcal{S}}}}{\left| \partial_q\ave{\hat{\mathcal{S}}} \right|^2} \label{delta}
\end{eqnarray}
where the variance of the signal $\hat{\mathcal{S}}$ is given by $\Delta^2{\hat{\mathcal{S}}}=\ave{\hat{\mathcal{S}}^2}-\ave{\hat{\mathcal{S}}}^2$. Not always the signal saturates the upper bound of sensitivity ({\ref{QCRB}}). Nevertheless, it has the advantage of being easier to be measured in a realistic experiments. On the other hand, having access to the error-propagation formula (\ref{delta}) only needs the first and second moments of the signal (i. e. $\ave{\hat{\mathcal{S}}}$ and $\ave{\hat{\mathcal{S}}^2}$ respectively).

The upper bound for the scaling of the above introduced QFI is set by critical exponents for the second-order quantum phase transition. It was shown that $F_q\propto N^{\mu}$, where $\mu=2/(d\nu)$ and $\nu$ as the critical exponent satisfying the divergence of correlation length and $d$ is the effective spatial dimension, as explained in~\cite{Rams2018,Pezze2019}. 
No bound is expected for the scaling exponent of the QFI around first-order quantum phase transition. On the other hand, the standard quantum limit is equal to the total atoms number $N$ for the zero temperature case. In addition, we point out that the QFI in \eqref{QFI_def}, where the states entering in the fidelity are separated by a variation of $q$, is a different quantity from that defined in quantum interferometry where states are linked by a unitary operation~\cite{Pezze2014_rev}. Notably, only the QFI defined in the second way, $G$, is a witnesses of the multipartite entanglement and subject to the Heisenberg limit for its scaling with the number of particles $N$ , i.e. $G(N) \sim N^{\zeta}$, $\zeta \leq 2$~\cite{Pezze2009_Heis, Pezze2014_rev}.

\section{Precise estimation around criticality} \label{sec3}

The antiferromagnetic spin-1 system exhibits two different types of criticalities, depending on the value of the magnetization, and characterized by different behaviours of the second derivative of the ground state energy, as mentioned in Section \ref{sec1}. In the following, we exploit quantum criticalities of the spin-1 antiferromagnetic condensate in transverse magnetic field 
to demonstrate high sensitivity in estimation of the coupling constant $q$ using the QFI formalism introduced above. We also provide the useful experimental signal $\hat{\mathcal{S}}$ which almost saturates the QCRB.

\subsection{Zero magnetization} \label{sec3a}

\begin{figure}[]
	\centering {\includegraphics[width=\columnwidth]{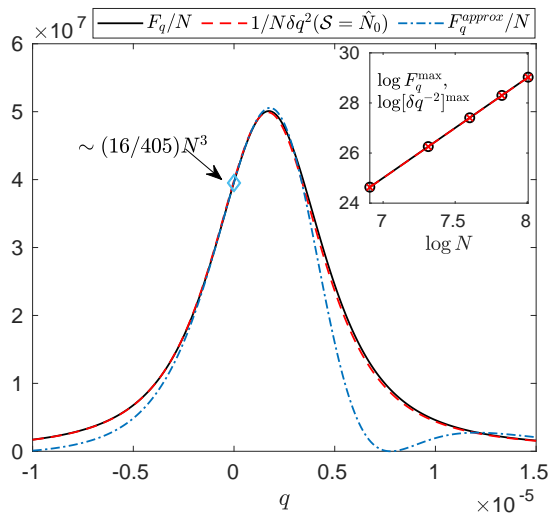}} 
	\caption{(Color online) The scaled quantum Fisher information $F_q/N$ (the solid black line), the precision $1/N\delta q^2$ (dashed red) and the analytical $F_q^{approx}/N$ results up to second correction using perturbation theory around $q \sim 0$ (dash-dotted blue) versus the parameter $q$ for $N=1000$ and $M/N=0$. The turquoise diamond marks the explicit analytical value of the QFI at $q=0$, i.e. $F_q(q=0)/N=(16/405)N^3 $. Note that the perturbation theory prediction is in good agreement with the $F_q$ in the validity range of the perturbative approach, that is for small enough $q$. In the inset we give the logarithm of maxima for $F_q$ and $1/\delta q^2$ ($\hat{\mathcal{S}}=\hat{N}_0$) depending on $\log N$.}  
	\label{fig:fig3}
\end{figure}

Let us consider first the case of $M=0$ where the system shows a discontinuous QPT in thermodynamic limit with the critical point at $q_c=0$~\cite{Zhang2003}.
The variation of $F_q$, defined as (\ref{QFI}), versus $q$ for $N=1000$ is shown in Fig.~\ref{fig:fig3}. Obviously, the value of $F_q$ increases significantly around criticality dropping down far away from the critical region. We also show in the same figure the sensitivity estimated by the error-propagation formula (\ref{delta}) when the signal is set to the number of atoms in the $m_f=0$ Zeeman state, $\hat{\mathcal{S}}=\hat{N_0}$. The inverse of (\ref{delta}) almost saturates the QFI, and we observe the same behavior for both quantities when increasing the precision around criticality. Note that the maxima of $F_q$ and $1/\delta q^2$ are shifted to the same extent with respect to the mean-field critical point $q_c=0$ due to the finite number of atoms considered. That is by increasing $N$, the evaluated value of $q_c$ for a finite system approaches the prediction of the mean-field formalism for $N\rightarrow \infty$~\cite{Zhang2003}. 

In order to see how the total number of atoms affects the precision in the inset of Fig.~\ref{fig:fig3}, we show the logarithms of the maxima of $F_q$ and  $1/\delta q^2(\hat{N}_0)$ versus $\log{N}$. Both of them exhibit exponential behavior. The fitting gives the same scaling exponents for the QFI and for the inverse $1/\delta q^2$, i.e.~$F_q^{\max}\sim [1/\delta q^2]^{\max} \sim 0.05 N^{4}$ which beats the sub-shot-noise sensitivity $\sim N$. 
The identification of the particular signal that saturates the QFI is important from the 
experimental point of view. The reason is that in order to find the QFI, one needs to make the full tomography of the density matrix and subsequently to evaluate the QFI (\ref{fidel}) which is hardy possible for large systems, as mentioned before. Alternatively, one may extract the classical Fisher information, i.e. $F_c=\sum_x {\left(\partial_q P(x|q)\right)^2}/{P(x|q)}$, where $P(x|q)$ is the 
probability distribution of getting $q$ conditioned on making measurements over all eigenvalues of an observable $\hat{X}$ ($\hat{X}|x\rangle=x|x\rangle$)~\cite{Strobel:2014}. The optimization of the classical Fisher information over all possible observables approaches the QCRB marked by the QFI. On the other hand, the measurement of the number of atoms in the $m_f=0$ Zeeman component and extracting its first and second moments is more easily accessible in practice and yet provides essentially the same information. 

As the region of criticality is around $q=0$, when the magnetization is zero, it is convenient to use the perturbation theory to approximate the eigenvalues and eigenvectors of the Hamiltonian (\ref{H}). Using them, it is possible to approximate the QFI value around criticality and to extract the corresponding scaling exponent. To this end, we employ the second-order perturbation theory formalism for small values of $q$. Suppose that the unperturbed Schr\"odigner equation (with $q= 0$) has $\hat{H}_0|\psi_n^{(0)}\rangle=E_n^{(0)}|\psi_n^{(0)}\rangle$. When the parameter $q$ is small but nonzero, the idea is to express the Schr\"odinger equation $(\hat{H}_0+q\hat{H}_q)|\tilde{\psi}_n\rangle=\tilde{E}_n|\tilde{\psi}_n\rangle$ up to the second order corrections
\begin{eqnarray}\nonumber
\tilde{E}_n&=&{E}_n^{(0)}+q{E}_n^{(1)}+q^2{E}_n^{(2)}+\mathcal{O}(q^3)\\
|\tilde{\psi}_n\rangle&=&|\psi_n^{(0)}\rangle+q|\psi_n^{(1)}\rangle+q^2|\psi_n^{(2)}\rangle+\mathcal{O}(q^3),
\end{eqnarray}
where every single term can be expressed by the eigenvalues and eigenstates of unperturbed Hamiltonian $\hat{H}_0$~\cite{Sakurai1994}.
Consequently, one can find the explicit corrections to eigenenergies $\tilde{E}_n^{(1,2)}$ and eigenstates $|{\psi}_n^{(1,2)}\rangle$ of the spin-1 system when $\hat{H}_0=\frac{c}{2N}\hat{J}^2$ and $\hat{H}_q=-\hat{N}_0$. More details about the derivation of the results are presented in Appendix~\ref{app3}~(see equations (\ref{p_gs}), (\ref{P_S}) and (\ref{P_V})). 
In Fig.~\ref{fig:fig3} we show the QFI calculated using the expression (\ref{QFI}) and approximated eigenvalues and eigenstates. The agreement between exact results and approximated by the perturbation theory is found when $q\sim 0$.
The value of the QFI can be calculated analytically at $q=0$ using the approximated eigenstates and eigenvectors (\ref{p_gs}), (\ref{P_S}) and (\ref{P_V}). We found out that $F_q(q=0) \approx  N^3(N+3) 16/405$, which is $\propto 0.04N^4$. This result is in excellent  agreement with the numerical prediction for the QFI value as demonstrated in Fig.~\ref{fig:fig3}. To this purpose, one may also consider the ratio $1/\delta q^2$ with the signal $\hat{\mathcal{S}}=\hat{N}_0$ by inserting the approximated ground state (\ref{p_gs}) in (\ref{delta}). We obtain $1/\delta q^2=\frac{16}{405}N^4$ for small $q$ using the first- and second-order moments of $\hat{N}_0$ and (\ref{N02}). This result is in the complete agreement with the approximated QFI value.

 Finally, we notice that for a second-order quantum phase transition an algebraic scaling for the QFI is expected \cite{Hauke2016}, similarly as for the relevant observables (see e.g. \cite{cardy1996}). This behaviour is implied by scale invariance, in turn allowed by the vanishing of the mass gap in the thermodynamic limit. On the contrary, at a first-order quantum phase transition, the same scalings can occur only provided that the correlation length (of the order of the inverse of the mass gap) is at least equal to the finite size of the considered system. A similar situation occurs for instance for the two points connected correlations. For both the types of quantum phase transition, algebraic ansatzs for the scalings of the observables can be adopted for finite-size analysis.

\subsection{Macroscopic magnetization}\label{sec3b}

\begin{figure}[]
	\centering {\includegraphics[width=\columnwidth]{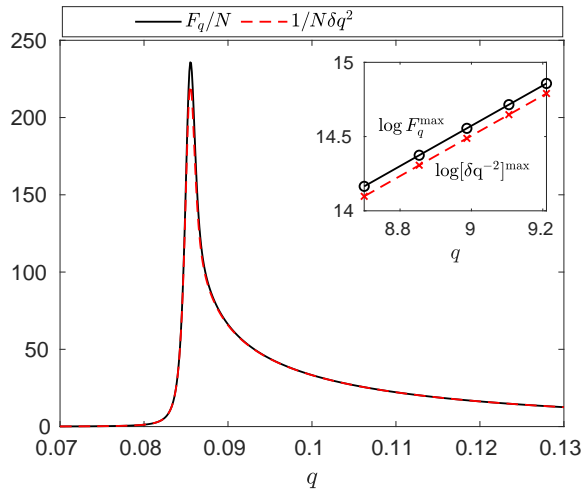}} 
	\caption{(Color online) The scaled quantum Fisher Information $F_q/N$ (the solid black line) and the scaled precision  $1/(N\delta q^2)$ calculated with $\hat{\mathcal{S}}=\hat{N}_0$ (dashed red line) versus $q$ for $N=6000$ and $M/N=0.4$. Here, the critical point is shifted compared to the one given by mean-field at $q_c=0.084$. {Inset: we give the logarithm of maxima for $F_q$ and $1/\delta q^2$ ($\hat{\mathcal{S}}=\hat{N}_0$) versus $N$. }}
	\label{fig:fig4}
\end{figure}

When the magnetization is non-zero, a continuous quantum phase transition occurs in the system as discussed in Section \ref{sec2}. In this case, the position of the critical point is $q_c=1-\sqrt{1-(M/N)^2}$ as shown by the mean-field approach~\cite{Zhang2003}. The variation of the QFI and of the inverse of signal-to-noise ratio for  $\hat{\mathcal{S}}=\hat{N}_0$ versus $q$ are shown in Fig.~\ref{fig:fig4} using $N=6000$ and $M/N=0.4$. Similarly as in the case of zero magnetization, the values of $F_q$ and $1/\delta q^2$ increase significantly around criticality. We extracted the scaling of the maxima of the QFI and of $1/\delta q^2$. We show its logarithms versus $\log N$ in the inset of Fig.~\ref{fig:fig4}. We obtained  $F_q^{\max}\sim 10.7 N^{1.36}$ and $[1 /\delta{q}^2]^{\max}\sim 10.05N^{1.36}$, which have the same power law scaling versus $N$ with different pre-factors. 
In the case of macroscopic magnetization we observe $F_q\sim N^{4/3}$. It means that $\mu = 4/3$.

The agreement between the scaling exponents of the QFI and of $1/\delta q^2$  has been generally demonstrated to occur at second-order quantum phase transitions, provided that the signal coincides with the order parameter ~\cite{Pezze2019}. In the same paper this equivalence has been measured explicitly for a bosonic Josephson junction model realized in a ultracold atoms set-up. It is interesting that the same agreement arises in our model at the first-order transition, perhaps due to the appearance of a scaling regime at the considered finite sizes. It is also worth to notice that the scaling exponent that we found here gives the same scaling as for the QFI (or equivalently fidelity susceptibility) in the Lipkin-Meshkov-Glick~\cite{Kwok2008}, Dicke~\cite{Liu2009} and bosonic Josephson junction~\cite{Pezze2019} models.

\begin{figure}[]
	\centering {\includegraphics[width=\columnwidth]{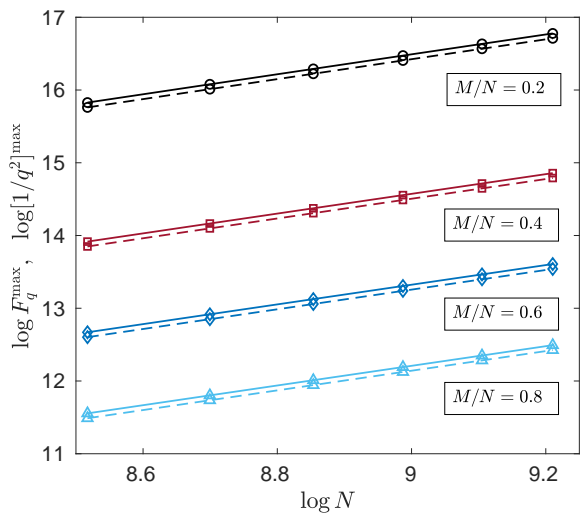}} 
	\caption{(Color online) The logarithm of the maxima for $F_q$ (solid lines) and for  $1/\delta q^2$ (dashed lines) versus $\log N$ for different values of $M/N=0.2,\,0.4,\,0.6$ and $0.8$. The values of both maxima increase with the number of particles $N$ and follow the power law ${F_q^{\max}} \propto N^\mu$ ($[1/\delta q^2]^{\max}\propto N^{\mu'}$ ) with fixed $\mu$ ($\mu'$).}
	\label{fig:varyM}
\end{figure}

\begin{table}[]
\begin{tabular}{ |p{1.5 cm}||p{1.5cm}|p{1.5cm}|p{1.5cm}|p{1.5cm}|  }
	\hline
	$M/N$ & $0.2$ & $0.4$& $0.6$ & $0.8$ \\
	\hline
	${\mu}$   & 1.37    & 1.36&   1.35 & 1.35 \\
	${\mu'}$&   1.37  & 1.36   & 1.35 & 1.35\\
	${\alpha}$&   -0.35  & -0.34   & -0.34 & -0.34\\
	\hline
\end{tabular}
\caption{The values of the scaling exponent for the maxima of the quantum Fisher information $F_q^{max}\propto N^{\mu}$, for the inverse of the error-propagation formula $[1/\delta q]^ {max}\propto N^{\mu'}$, and for the energy gap $\Delta_N^{min}\propto N^{\alpha}$ versus the fractional magnetization $M/N$. Here, the values are extracted by fitting to the numerical data.}
\label{tab1}
\end{table}

In order to demonstrate how increasing the magnetization changes the estimation precision, in Fig. \ref{fig:varyM} we show the maximum values for the QFI and for $1/\delta q^2$,
versus $\log N$ for different values of $M/N$. The maximum value of $\log F_q$ grows by increasing the number of particles for different values of the fractional magnetization. Moreover, by increasing $M/N$, the maximal values of the QFI  decrease while they still display very similar slopes. This suggests that the scaling of the QFI with $N$ can have the same scaling power law with the same scaling exponent: 
\begin{equation}\label{Fmax}
F_q^{max}\propto N^{\mu}. 
\end{equation}
To be more precise, in Table \ref{tab1} we provide the 
scaling exponents for each of the line shown in Fig.~\ref{fig:varyM} versus fractional magnetization $M/N$. The value of $\mu$ is almost fixed for different values of the fractional magnetization $M/N$. In addition, a similar behavior is observed for $1/\delta q^2$, $[1/\delta q^2]^{max} \propto N^{\mu'}$ \footnote{{The scaling properties can analyzed in terms of critical exponents such that $\mu=2/(d\nu)$ with $\nu$ as the critical exponent satisfying the divergence of correlation length and $d$ as the effective spatial dimension~\cite{Rams2018,Pezze2019}.}}. 
We will discuss this point in more details in Sec. \ref{conclusion}.

Up to now, we have considered the effect of criticality in the ideal case of zero temperature. Nevertheless, in the realistic situations  the temperature is always above the absolute zero and the system (\ref{H}) is never  in a pure rather in mixed states. Motivated by this fact, in the following section we consider the effect of a finite temperature on the estimation precision.

\section{The role of non-zero temperature}\label{sec4}

\begin{figure}[]
\begin{picture}(0,220)
\put(-125,120){\includegraphics[width=0.25\textwidth]{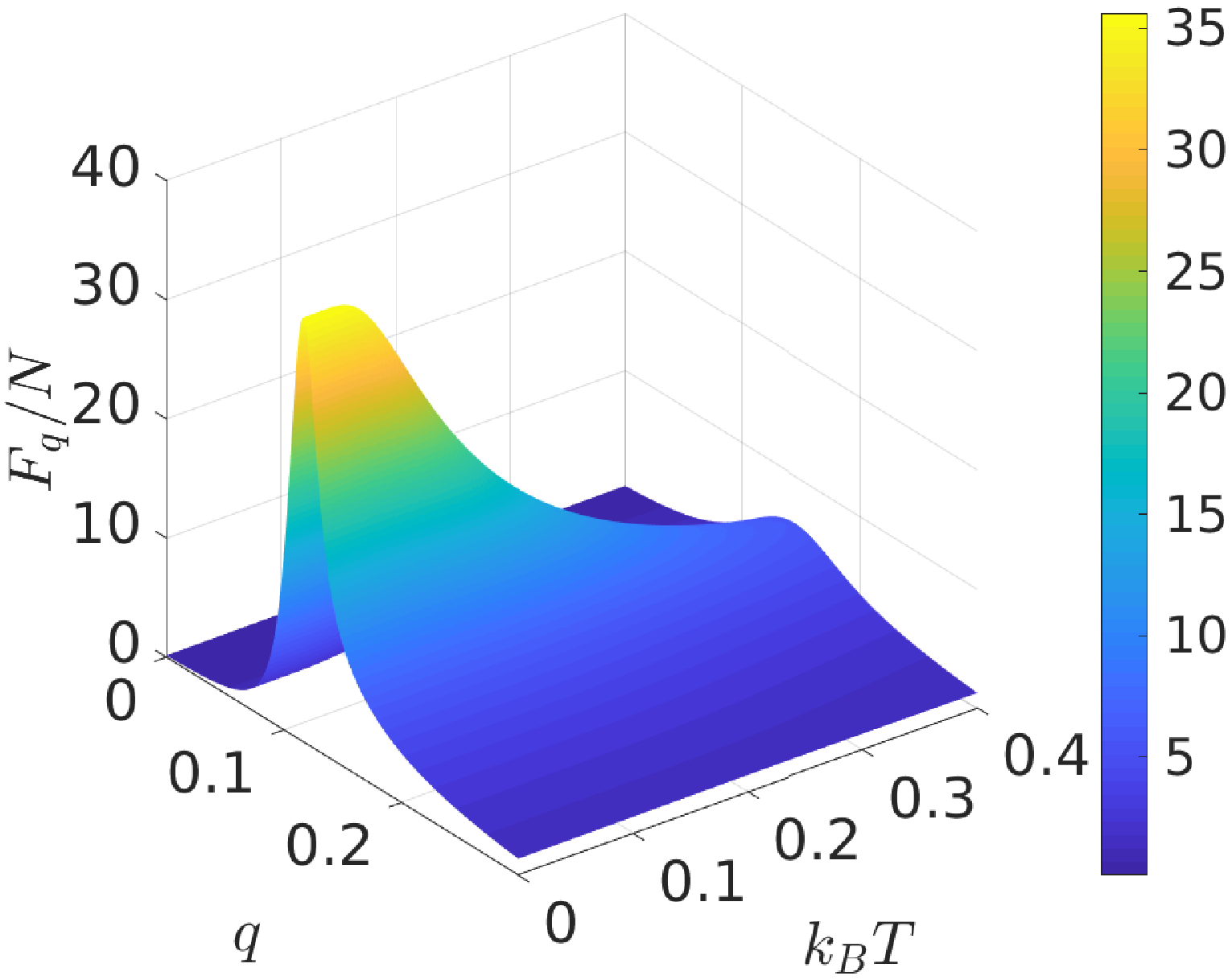}}
\put(-120,218){(a)}
\put(2,120){\includegraphics[width=0.25\textwidth]{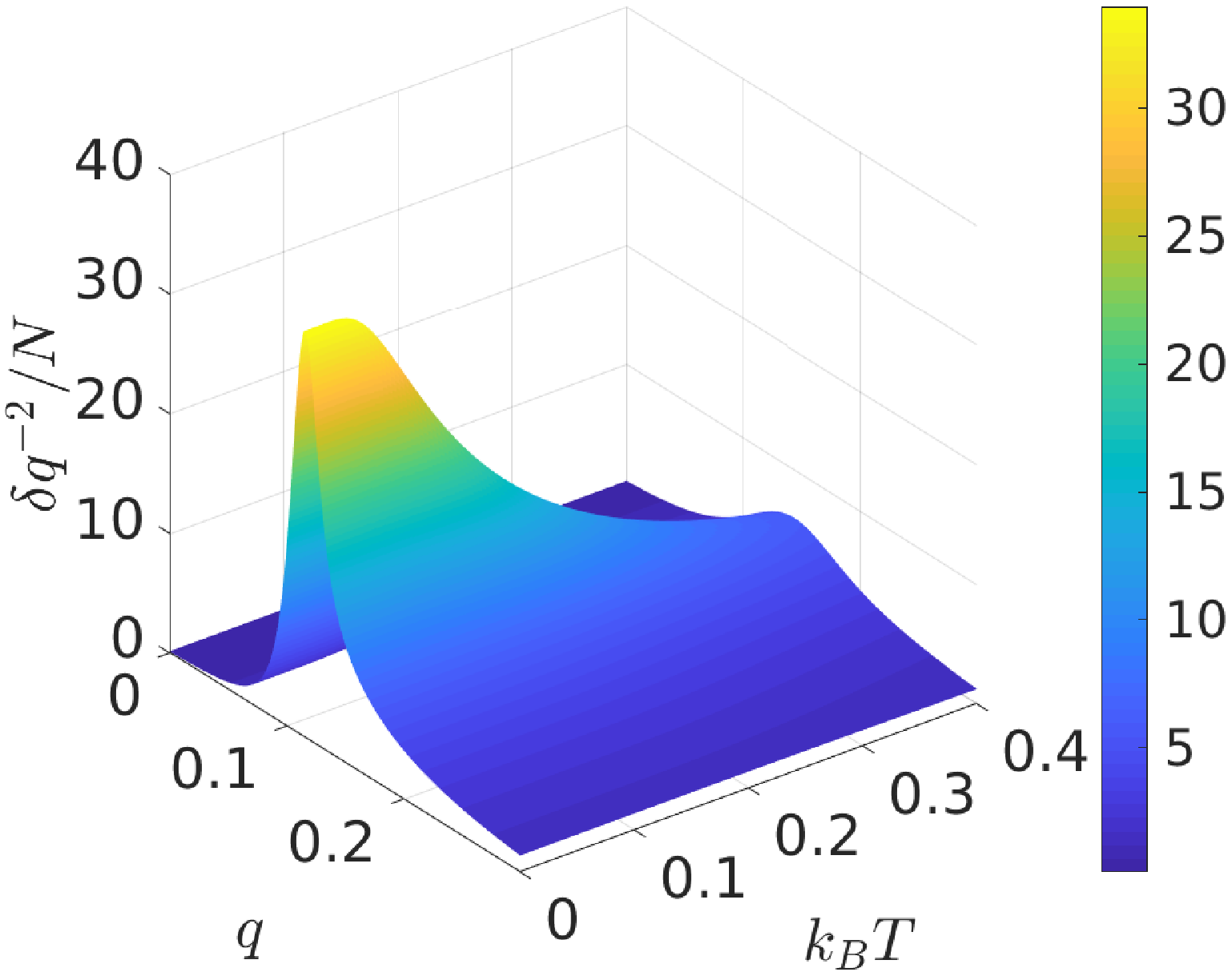}}
\put(7,218){(b)}
\put(-125,0){\includegraphics[width=0.25\textwidth]{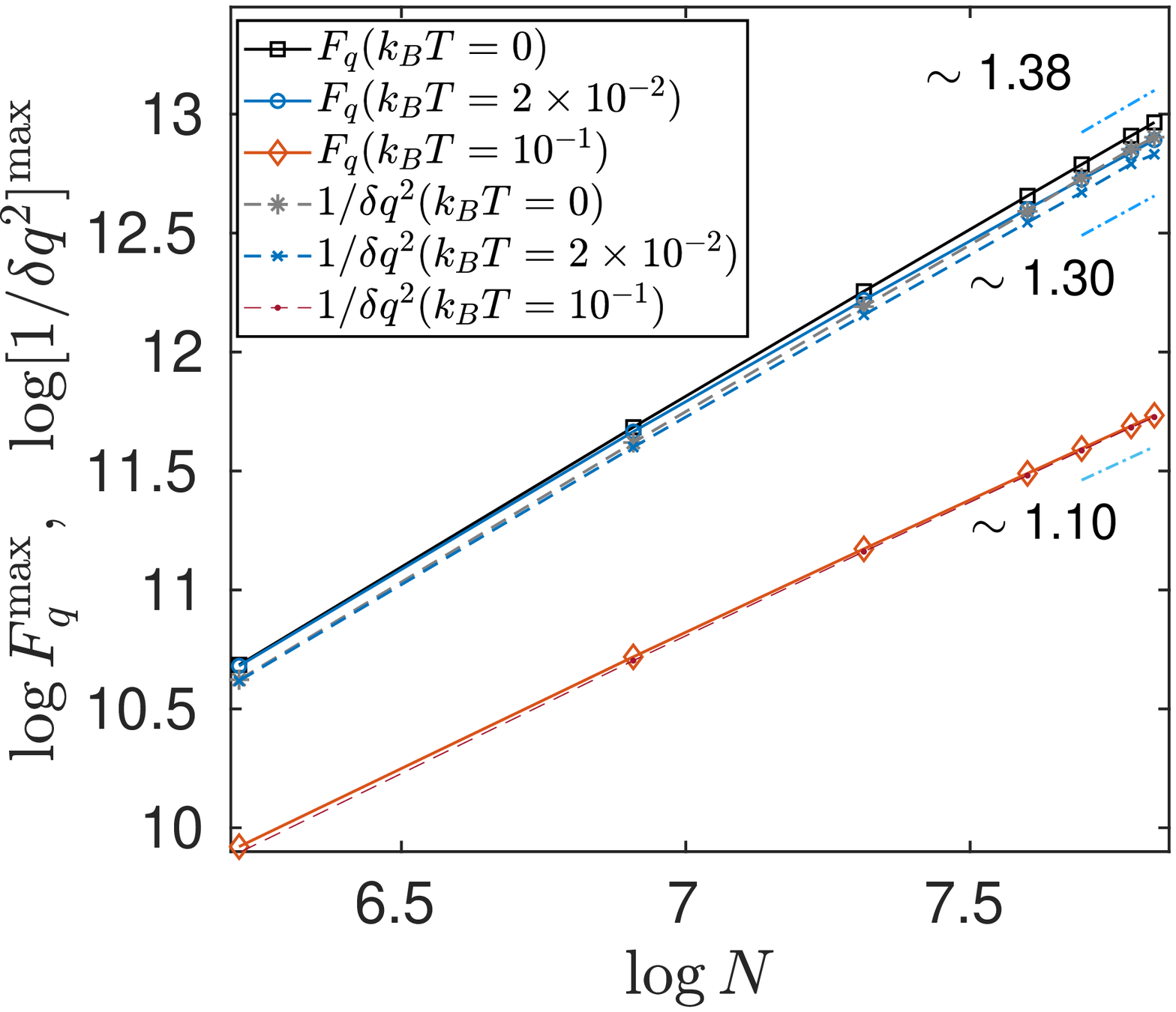}}
\put(-120,108){(c)}
\put(2,0){\includegraphics[width=0.25\textwidth]{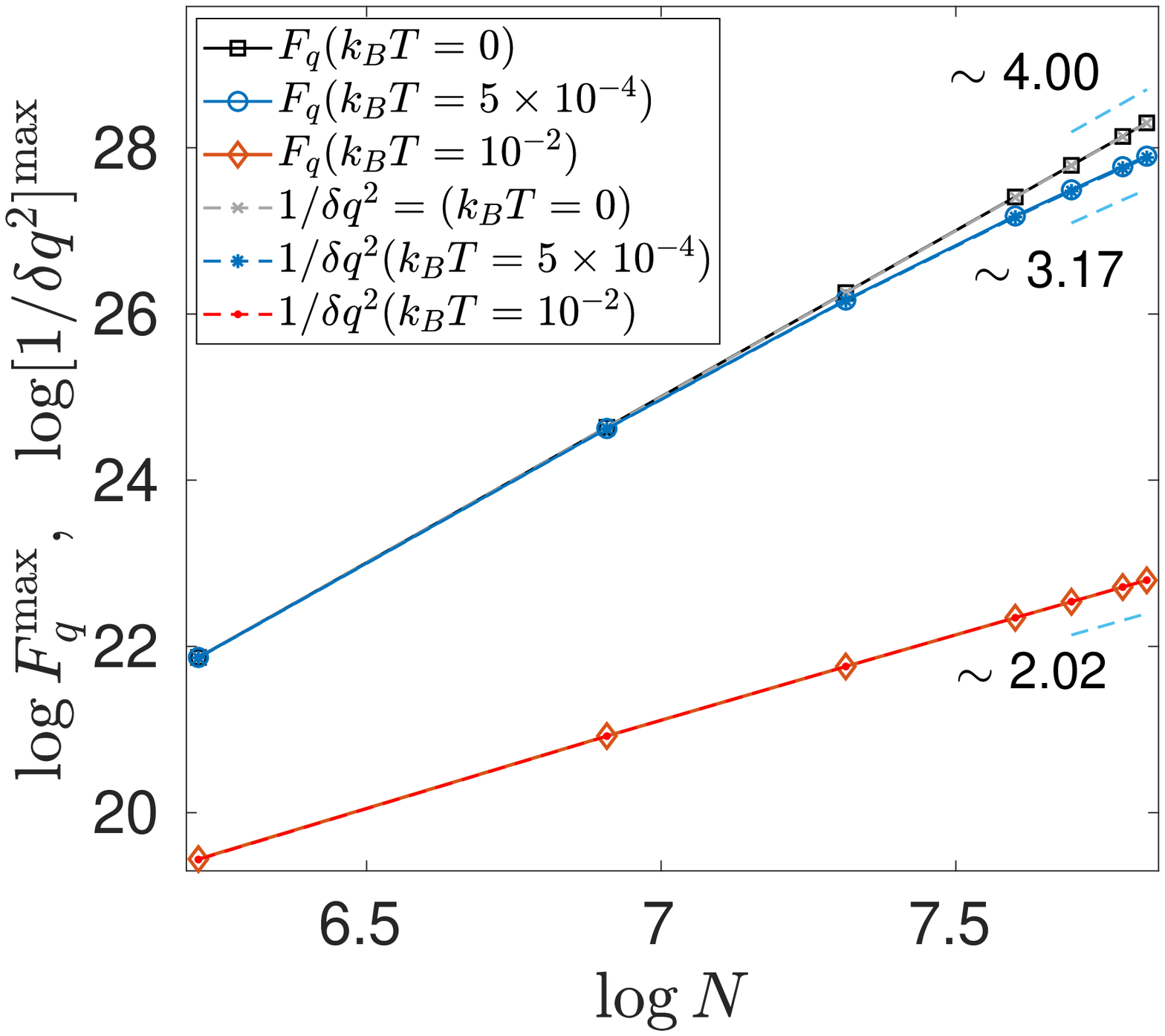}}
\put(7,108){(d)}
\end{picture}
\caption{The precision in estimation of the parameter $q$ quantified by $F_q$ (a) and the error-propagation formula through $1/ \delta q^2$ (b) versus $q$ and $k_B T$ for $N=100$ and $M/N=0.4$. The estimation precision reduces by increasing the temperature. This is due to rising the thermal fluctuations which beats the quantum effects. 
Logarithms of maximal values for $F_q$ (solid lines) and $1/\delta q^2$ (dashed lines) versus $\log N$ for $k_BT/c=0,\, 2\times 10^{-2}, \, 10^{-1}$
when $M/N=0.4$ (c) and $k_BT/c=0,\, 5\times 10^{-4}, \, 10^{-2}$ when $M/N=0$ (d). The sensitivity of estimating $q$ approaches the standard quantum limit for both the QFI and $1/\delta q^2$ when the temperature grows. In the case of zero magnetization (d), a change in the value of scaling exponent is observed for the lowest temperature ($k_BT/c=5\times 10^{-4}$). A few points on the right-hand side of the blue curve lie in the intermediate regime, where the scaling exponent is modified due to non zero temperature, because their corresponding energy gap are of the order of temperature.}
\label{fig:fig6}
\end{figure}

In the case of a non-zero temperature $T$, the quantum states of the system are described by the canonical Gibbs density matrix 
\begin{eqnarray}
\rho(q)=\sum_n \frac{e^{-E_n(q)/k_BT}}{Z} |\psi_n\rangle \langle \psi_n|
\end{eqnarray}
where the eigenstates are weighted by $w_n:={e^{-E_n(q)/k_BT}}/{Z}$ and $Z:={\rm tr}\left(e^{-H/k_B T}\right)$ is the partition function with the Boltzmann constant $k_B$. The QFI and the signal-to-noise ratio can be extracted using the equations (\ref{fidel}) and (\ref{fid}) \footnote{Alternatively, QFI can be derived using (\ref{fidel-p}).}. We focus on the case of macroscopic magnetizations at the moment and will return to zero magnetization later on. In Fig.~\ref{fig:fig6}, we provide the density plots of $F_q$ (a) and $1/\delta q^2$ (b) versus $q$ and $k_BT$ for $N=100$ and $M/N=0.4$. As the temperature is increased, the maximum value of $F_q$ and $1/\delta q^2$ approaches zero. For the case of zero magnetization, we have obtained the same qualitative plots as for macroscopic magnetization, given in Fig. \ref{fig:fig6}. However, the region that the QFI and $1/\delta q^2$ does not affect by temperature, pushes toward the lower temperature range. 

\begin{figure}[]
\centering
\begin{picture}(0,110)
\put(-125,0)
{\includegraphics[width=0.5\linewidth]{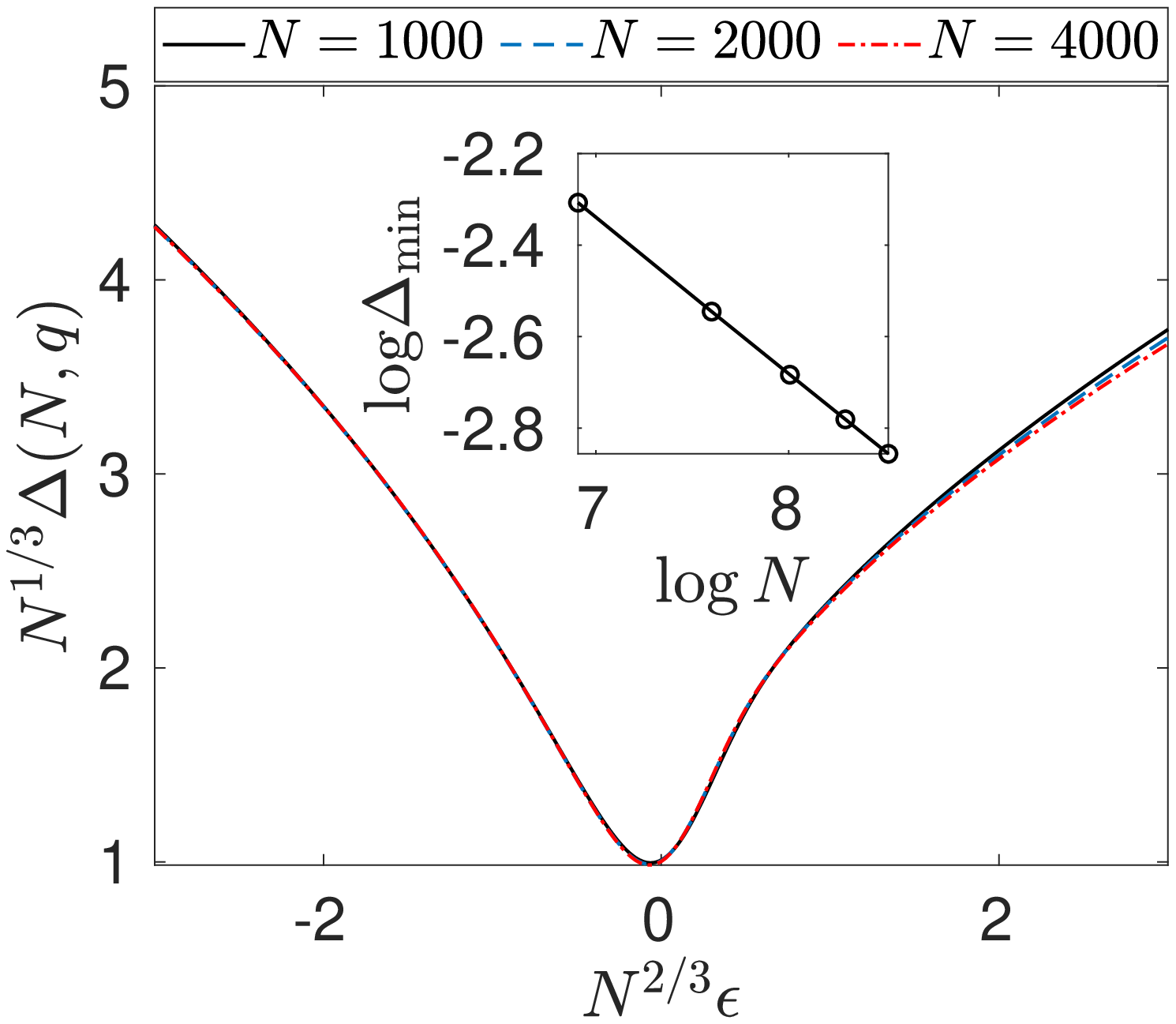}}
\put(-120,110){(a)}
\put(5,0)
{\includegraphics[width=0.5\linewidth]{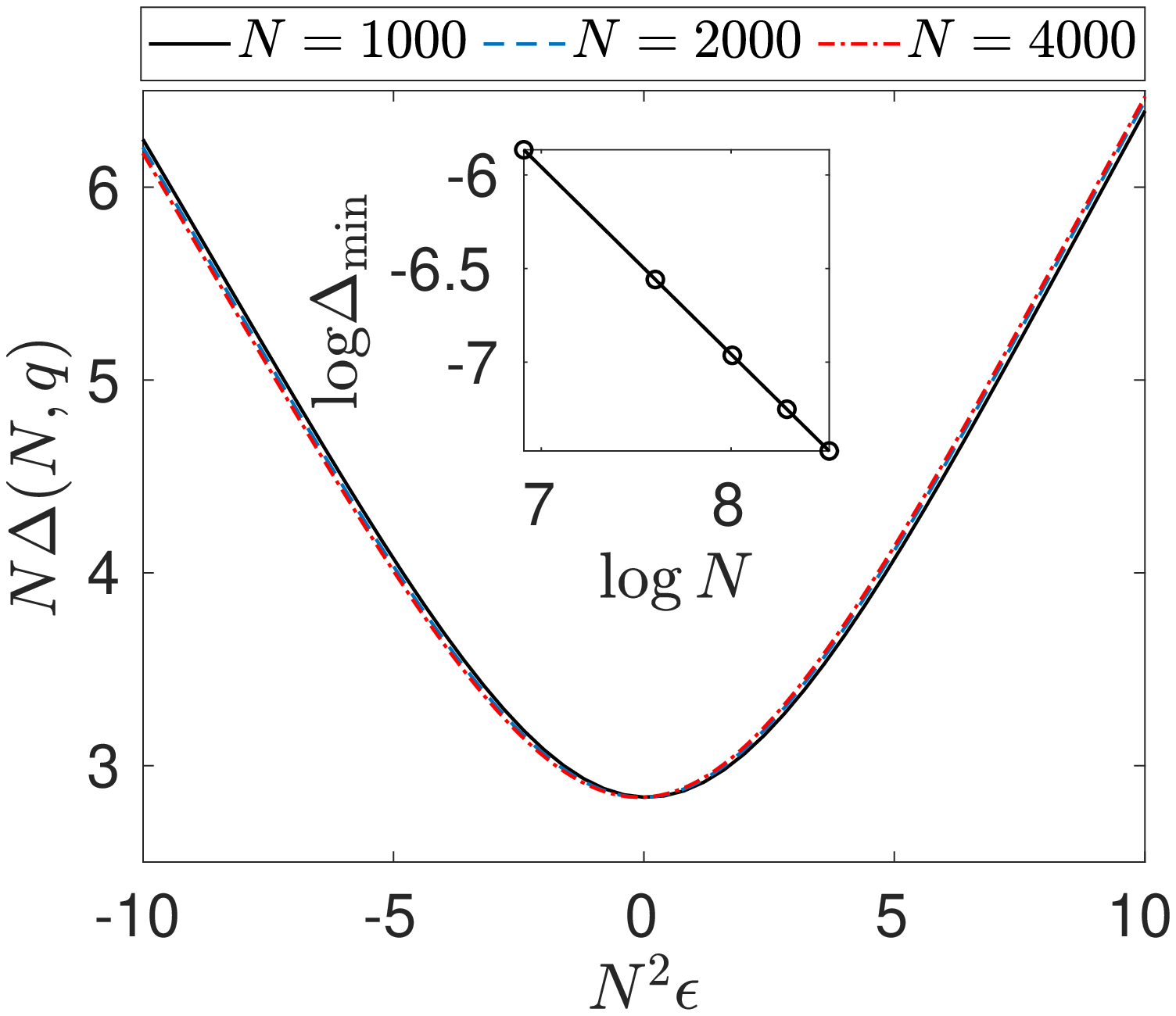}}
\put(25,110){(b)}
\end{picture}
\caption{(Color online) Scaling of the energy gap (\ref{gap2}) for $M/N=0.4$ (a), and $M=0$ (b). In the inset, the logarithms of the energy gap minimum are shown versus $\log N$. The fits confirm the power law behaviour (\ref{gap1}) with the scaling exponents $\alpha={-0.34}$ for $M/N=0.4$, and $\alpha = {-1}$ for $M=0$.}
\label{fig:fig8}
\end{figure}

In order to understand the finite temperature behavior, let us first investigate the energy gap $\Delta(q, N)=E_0(q)-E_1(q)$, defined as the energy difference between the ground state and the first excited state of the Hamiltonian (\ref{H}), close the criticality. The minimum of the energy gap is expected to be subject to the asymptotic law~\cite{Vidal2006,Vidal2004}
\begin{equation}\label{gap1}
\Delta_{\rm min} \propto N^{\alpha},
\end{equation}
while its variation to a scaling function of the form
\begin{equation}\label{gap2}
\Delta(N,q)=\Delta_{\rm min}\, f\left( N^{\beta}\epsilon \right),
\end{equation}
where $f(x)$ is the homogeneous function and $\epsilon = {q-q^N_c}$ with $q^N_c$ being a position of the energy gap minimum. The scaling exponents $\alpha$ and $\beta$ are independent of the system size, and moreover they are the same for all systems belonging to the same universality class. We verified the energy gap scaling and we demonstrate it for $M/N=0.4$ in Fig.~\ref{fig:fig8}(a). 
In addition, in the inset of Fig.~\ref{fig:fig8}(a), we provide the values of fitted scaling exponent, and show that it scales as $\Delta_{\rm min}\propto N^{-0.34}$ for macroscopic magnetization $M/N=0.4$. In table \ref{tab1}, we give the scaling exponent of the energy gap minimum $\alpha$ versus other values of the fractional magnetization $M/N$, all of them are close to ${-1/3}$. Our findings are consistent with the prediction for antiferromagnetic condensate~\cite{dressing2014}. Moreover, the scaling exponent of the energy gap is the same as that for a ferromagnetic spinor condensate \cite{Duan2013}, the Lipkin-Meshkov-Glick ~\cite{Vidal2004}, bosonic Josephson junction~\cite{Pezze2019} and the interacting Dicke   \cite{Vidal2006} models. In the case of zero magnetization the energy gap minimum scales as $N^{-1}$~\cite{Gerbier2016,dressing2014,Sarlo2013}. Although the universal behavior cannot be expressed in terms of critical exponents, it indeed can be observed in Fig.~\ref{fig:fig8}(b) with the scaling of the energy gap minimum as~$\Delta_{\rm min}=2.83N^{-1}$.

\begin{figure}[]
\centering
\begin{picture}(0,110)
\put(-125,0)
{\includegraphics[width=0.5\linewidth]{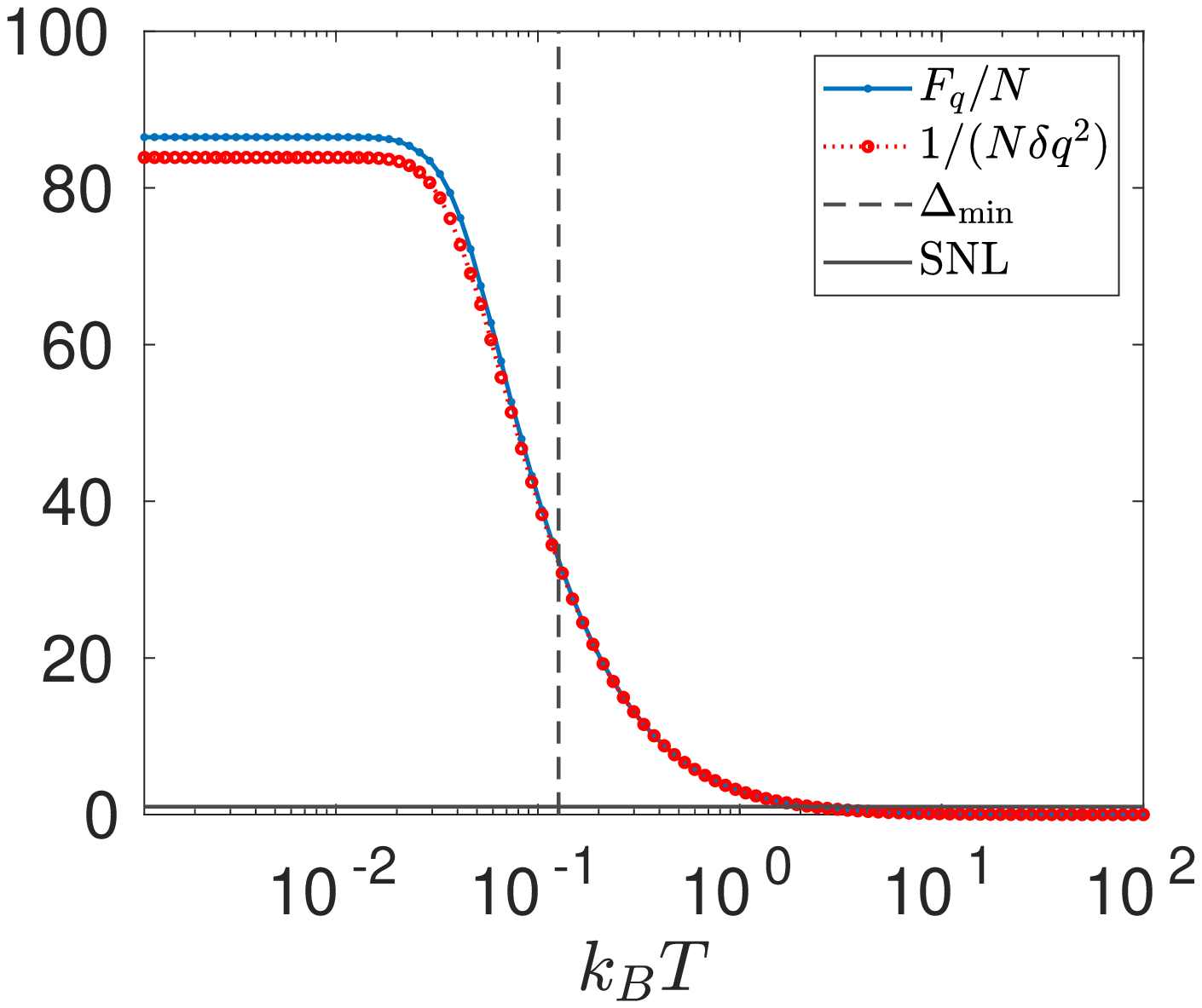}}
\put(-120,105){(a)}
\put(5,0)
{\includegraphics[width=0.5\linewidth]{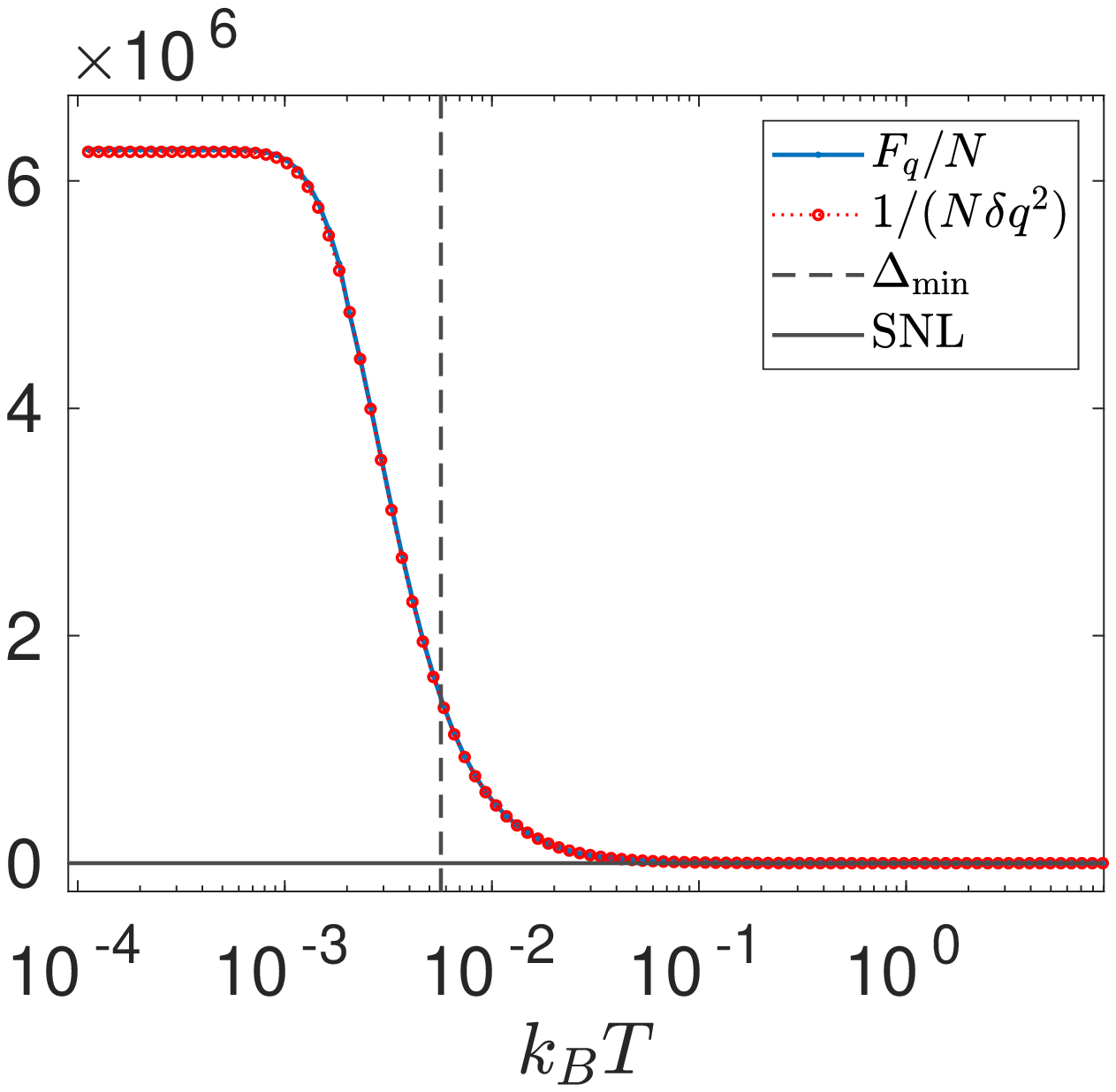}}
\put(25,105){(b)}
\end{picture}
\caption{(Color online) Decay of the maximal values of QFI and of $1/\delta q^2$ in the presence of a non-zero temperature $k_BT$ for $N=500$ and (a) $M/N=0.4$ and (b) $M=0$. The vertical dashed line marks the value of the energy gap while the solid black line marks the zero temperature limit. When the temperature is of the order of the energy gap, one can still obtain the sub-shot-noise sensitivity.}
\label{fig:fig9}
\end{figure}

Having explored the energy gap of the system, we can now achieve a better understanding of the finite temperature behavior of both $F_q$ and $1/\delta q^2$. We demonstrate that in Fig.~\ref{fig:fig9} for the macroscopic $M=0.4$ and zero magnetization in the panels (a) and (b), respectively. Three different regimes of the QFI (and $1/\delta q^2$) can be distinguished depending on the temperature value compared to the energy gap: i) the quantum (zero-temperature) regime for $k_BT \ll \Delta_{\rm min}$, ii) the intermediate one, when $k_BT\approx \Delta_{\rm min}$ and iii) the classical one for $k_BT\gg\Delta_{\rm min}$ \cite{Gabbrielli2018,Gabbrielli2019,sachdev2011}. In the first regime, the $F_q$ and $1/\delta q^2$ are robust against thermal fluctuations. However, as the temperature increases, both the QFI and $1/\delta q^2$ decreases. In the third regime, the scaling of the QFI approaches the classical shot-noise limit (SNL), which is $\sim N$. Moreover, it is interesting to note that for the case of first-order phase transition the quantum robust regime is pushed toward the lower temperatures, compared to that of the second-order phase transition. This is due to the fact that in the later case, the (finite-size) energy gap is three times smaller than the one for the second- order~\cite{dressing2014}. This squeezes the quantum robust regime to the lower temperatures in the case of the zero magnetization.

To investigate the effect of a finite temperature more quantitatively, in panels (c) and (d) in Fig.~\ref{fig:fig6} we show the corresponding logarithmic values of $F_q^{\max}$ and $[1/\delta q^2]^{\max}$ versus $\log N$ changing the temperature from $0$ to $k_BT/c=2 \times 10^{-2}, 10^{-1}$ for the magnetization $M/N=0.4$ (c) and to $k_BT/c=0.0005, 0.01$ for zero magnetization (d). 
As we see for the large enough values of $N$ and macroscopic magnetization, the scaling exponent for the maximum value of the QFI is reduced from ${1.38}$ for zero temperature to ${1.30}$ for ($k_BT=0.02$), and further to ${1.1}$  close to the shot-noise limit. 
A decrease of scaling exponent also occurs for zero magnetization when it changes through $3.17$ ($k_BT/c=5 \times 10^{-4}$) and $2.02$ ($k_BT/c=10^{-2}$) approaching the shot-noise limit for higher temperatures.

\vspace{2cm}

\section{Discussion and Conclusion }\label{conclusion}

In the previous sections, we have discussed  the effect of criticality in a spin-1 BEC located in a transverse magnetic field in order to estimate the value of coupling constant. We have discussed that the precision of estimation of $q$ depends on the type (nature) of criticality we employ.

To this end, we have made use of the quantum Fisher information as a theoretical criterion to estimate the sensitivity of our spinor sensors around the critical region. In addition, we have considered the sensitivity evaluated using the error-propagation formula. We introduced the respective signal (equivalent to the population in $m_f=0$ manifold, i.e. $\hat{\mathcal{S}}=\hat{N}_0$). The identification of this simple-to-measure signal and the error-propagation formula is of experimental importance as it contains the variance $\Delta^2\hat{N}_0$ and the average population $\ave{\hat{N}_0}$, which makes it possible to find the sensitivity much more easily than by the QFI measurement. Indeed, evaluating the QFI requires the state tomography of the system density matrix $\hat{\rho}$ which might be a challenging task for the ensembles consist of thousand atoms. 

Firstly, we have shown that a first-order quantum phase transition is realized for the zero magnetization in the system, when the transition from the polar to the antiferromagnetic phases occurs. For finite-size spinor condensate with total number of atoms $N$, we have found that the QFI and inverse of the signal-to-noise ratio $1/\delta q^2$ scale $\propto N^4$. We also investigated the behaviour of the QFI around the transition between the antiferromagnetic and broken axisymmetry phases which occurs for macroscopic magnetization. In this case, we calculated the scaling of the QFI versus $N$ as $F_q\sim N^{4/3}$ around the critical point. We evaluated the same scaling factors for $1/\delta q^2$, finding the same qualitative behaviour as for the QFI. The reason for decreasing the sensitivity with increasing magnetization lies in the fact that the quantum Fisher information is related to the distinguishability of the quantum states of the different phases around the critical points. 

The scaling observed by us can also be analyzed in a different way.
As we see in Fig.~\ref{fig:fig2}, the ground state exhibits much more pronounced change around the criticality in the case of zero magnetization. In order to have a physical sense, we consider the error-propagation formula (\ref{delta}). In this regard, one can show that in zero magnetization case, where the first-order quantum phase transition occurs, both of the variance $\Delta^2 \hat{N}_0$ and the slope of the zero manifold population $\ave{\hat{N}_0}$, dependent on $q$, are maximized. In particular at the critical point, the variance of $\hat{N}_0$ scales $\propto N^2$ \cite{Gerbier2012}, while $\partial_q\ave{\hat{N}_0}\propto N^3$ (see Appendix A for explicit expressions). As a result,  $\delta q^2$ scale as $N^4$ around criticality. On the other hand, in the non-zero magnetization case, hosting the second-order phase transition, it is not the variance  $\Delta^2{\hat{N}_0}$ which is maximized around the critical point rather its slope $\partial_q ({\Delta^2\hat{N}_0})$ maximizes quite close to $q_c$. Nevertheless, the denominator $\partial_q \ave{\hat{N}_0}$ still increases around the QCP. As a result, in the second-order phase transition, the interplay of the nominator and denominator of (\ref{delta}) results in the scaling of $N^{4/3}$ for $\delta q^2$. In this case, the variance seems to change as $\Delta^2 \hat{N}_0\propto N^{2/3}$, while $\partial_q \ave{\hat{N}_0}\propto N$, which is less noticeable compared to the   scaling of the first-order transition, $\sim N^3$. This behaviour can be seen qualitatively from the slope of $\ave{\hat{N}_0}$ in Fig. \ref{fig:fig2}. 

\begin{figure}[]
	\centering {\includegraphics[width=1\columnwidth]{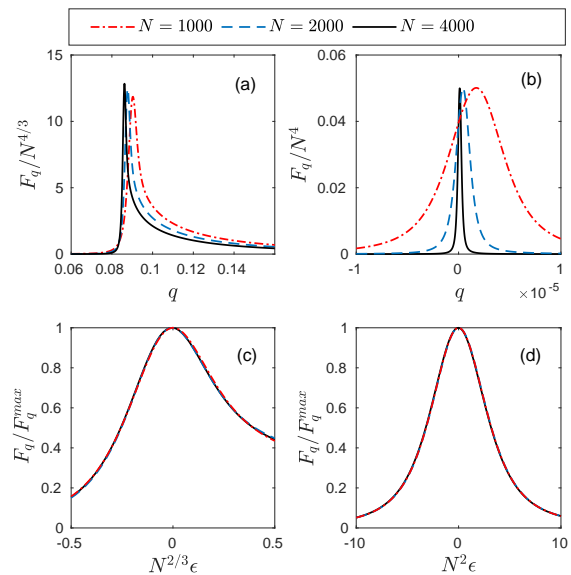}} 
	\caption{(Color online) Upper panel: scaling of the quantum Fisher information~(\ref{univ-FQ}) for $M/N=0.4$~(a) and $M=0$~(b), and for different values of the total atomic number $N$. Lower panel: universal behaviour of QFI versus dimensionless parameter $\epsilon=q-q_c^N$ for the same magnetization values.}
	\label{fig:fig11}
\end{figure}

It is worth to discuss also that the QFI is subject to the scaling hypothesis~\cite{Kwok2008,Liu2009}
\begin{equation}\label{univ-FQ}
\frac{F_q}{F_q^{\rm max}} = g(N^{\gamma}\epsilon ),
\end{equation}
with $F_q^{\rm max}$ being the maximum of the QFI (\ref{Fmax}) (corresponding to $\mu$), $\gamma$ representing the {scaling exponent} and $g(x)$ is a homogeneous function. For the sake of completeness, in Fig.~\ref{fig:fig11}(c) we show the scaling  of the QFI with $N$  (\ref{univ-FQ}) which is characterized by an exponents $\gamma=2/3$ and $\mu=4/3$ (\ref{Fmax}). The same scalings has been provided for the the QFI (or equivalently fidelity susceptibility) in the Lipkin-Meshkow-Glick~\cite{Kwok2008}, Dicke~\cite{Liu2009} and bosonic Josephson junction~\cite{Pezze2019} models~\footnote{In particular, in Refs. \cite{Kwok2008} and \cite{Liu2009} the fidelity susceptibility is given equivalently.}. Consequently, it suggests that the antiferromagnetic spinor condensate hosts a second-order quantum phase transition in the same universal class as these systems. It is interesting that the QFI for zero magnetization also seems to display a scaling in $N$, due to the fact that the correlation length is at least equal to the finite sizes of analyzed system (see Fig.\ref{fig:fig11}(d)). We have also provided the numerical results for the finite-size energy gap of our spinor system for zero and non-zero magnetizations, which turns out to be proportional to $N^{-1/3}$ and $N^{-1}$,  respectively {\cite{Gerbier2012}.} The finite size energy gap scales as that for other \emph{fully-connected} models, such as Lipkin-Meshkov-Glick \cite{Vidal2004,Heiss2005}, the Dicke
\cite{Vidal2006} and a ferromagnetic spinor condensate  with no spatial degrees of freedom \cite{Duan2013}.

In addition, we have taken into account the effect of non-zero temperature. Depending on the value of the energy gap compared to the temperature, different regimes of sensitivity appears. For low temperatures $k_B T\ll\Delta_{\rm min}$ the sensitivity witnessed by QFI and by $1/\Delta q^2$, is quite robust against thermal noise. increasing the temperature value reduces the sensitivity, until the limit $k_B T\gg\Delta_{\rm min}$, where the sensitivity highly diminishes and scales as the classical shot-noise sensitivity. We have found a qualitatively similar behaviour for both types of criticalities of the system. In the case of first-order transition, the sensitivity is much less robust against noise due to the smaller energy gap at finite sizes (Sec. \ref{sec4}).

For the experimental realization of our protocol, one possibility is to perform the adiabatic ramp of the ground state of the system followed by measuring the population of the atoms in $m_f=0$ Zeeman energy level, namely $\hat{N}_0$. The viability of these methods is connected to the energy gap~$\Delta$ as it determines the adiabatic evolution time $\tau$ according to the adiabatic criterion
$\hbar|\langle e |\dot{H}|g \rangle| \ll \Delta^2$~\cite{Gerbier2016,Comparat2009}, where $|g\rangle$ and $| e \rangle$ are the ground and excited states of the Hamiltonian \eqref{H}, respectively.
In the case of zero magnetization, the minimum of the energy gap between the ground and first excited state is $\sim N^{-1}$, while on the other hand $\langle e|\dot{H}|g\rangle \sim N/(3\tau)$ considering $\ave{\hat{N}_0}=N/3$ and a linear change in time of the parameter $q$. This gives $\tau \gg N^3/27$, which restricts the possibility of performing the adiabatic evolution of the ground state to relatively small systems. In the case of larger systems, it might be possible to use other methods, such as shortcut to adiabaticity discussed in \cite{Gerbier2016}. On the other hand, in the case of macroscopic magnetization, the process of adiabatic sweeping of $q$ is easier to be implemented due to the wider energy gap which scales as $N^{-1/3}$. In this case, it is possible to maintain adiabatic process using the microwave dressing, as discussed in Refs.~\cite{dressing2014,dressing2015}.  
A similar experimental work in the context of bosonic Josephson junction has been realized very recently \cite{Pezze2019}.

\section*{ACKNOWLEDGMENTS}
\noindent We gratefully acknowledge fruitful discussions with Matteo Paris, Luca Pezz{\`e}, Anna Sanpera and Augusto Smerzi. This work was supported by the Polish National Science Center Grants DEC-2015/18/E/ST2/00760 (SSM) and under QuantERA programme, Grant No. UMO-2019/32/Z/ST2/00016 (EW). This project has received funding from the QuantERA Programme under the acronym MAQS.

\appendix{}
\section{Collective spin-1 operators}\label{app1}

The matrix representation of the total spin 1 are defined as 
\begin{eqnarray}\nonumber
f_x&=&\frac{1}{\sqrt{2}}
\begin{bmatrix}
0  & 1 & 0 \\
1 & 0 & 1 \\
0  & 1 & 0  \\
\end{bmatrix}, \ \ 
f_y=\frac{i}{\sqrt{2}}
\begin{bmatrix}
0  & -1 & 0 \\
1 & 0 & -1 \\
0  & 1 & 0  \\
\end{bmatrix}, \\ \label{Fs}
f_z&=&
\begin{bmatrix}
1  & 0 & 0 \\
0 & 0 & 0 \\
0  & 0 & -1  \\
\end{bmatrix}.
\end{eqnarray}
In order to write the spin operator in terms of annihilation and creation operators let us start with the vector $\vec{\hat{\Psi}}^T=(\hat{\Psi}_{1},\hat{\Psi}_0,\hat{\Psi}_{-1})^T$, whose components under the single mode approximation are $\hat{\Psi}_{mf}(\vec{r})=\phi(\vec{r})\hat{a}_{mf}$ for $m_f=0,\pm 1$. If ${f_\alpha}_{ij}$ denotes the $(i,j)$th element of the $\alpha=x,y,z$ spin-1 matrix $f_{\alpha}$ and $\Psi_{i}$ the $i$-th element of the field operator, then the definition of collective spin operators $\hat{J}_{\alpha}=\Psi_{i}^{\dagger}[f_{\alpha}]_{ij}\Psi_j$ explicitly gives 
\begin{eqnarray}\nonumber
\hat{J}_{x}&=&\frac{1}{\sqrt{2}}(\hat{\Psi}_{-1}^{\dagger}\hat{\Psi}_{0}+\hat{\Psi}_{0}^{\dagger}\hat{\Psi}_{-1}+\hat{\Psi}_{0}^{\dagger}\hat{\Psi}_{+1}+\hat{\Psi}_{+1}^{\dagger}\hat{\Psi}_{0})\\ \nonumber
\hat{J}_{y}&=&\frac{i}{\sqrt{2}}(\hat{\Psi}_{-1}^{\dagger}\hat{\Psi}_{0}-\hat{\Psi}_{0}^{\dagger}\hat{\Psi}_{-1}+\hat{\Psi}_{0}^{\dagger}\hat{\Psi}_{+1}-\hat{\Psi}_{+1}^{\dagger}\hat{\Psi}_{0})\\ \nonumber
\hat{J}_{z}&=&\hat{\Psi}_{+1}^{\dagger}\hat{\Psi}_{+1}-\hat{\Psi}_{-1}^{\dagger}\hat{\Psi}_{-1}.
\end{eqnarray}
Subsequently, by replacing the field operator in terms of bosonic operators and considering the SMA following by integration over spatial degrees of freedom we get
\begin{align}\nonumber
	\hat{J}_{x} &\ =\ \frac{1}{\sqrt{2}}\left( \hat{a}^{\dag}_{\scriptscriptstyle{-1}}\hat{a}_{\scriptscriptstyle{0}} + \hat{a}^{\dag}_{\scriptscriptstyle{0}}\hat{a}_{\scriptscriptstyle{-1}} + \hat{a}^{\dag}_{\scriptscriptstyle{0}}\hat{a}_{\scriptscriptstyle{+1}} + \hat{a}^{\dag}_{\scriptscriptstyle{+1}}\hat{a}_{\scriptscriptstyle{0}}\right), \\
	\nonumber
	\hat{J}_{y} &\ =\ \frac{i}{\sqrt{2}}\left( \hat{a}^{\dag}_{\scriptscriptstyle{-1}}\hat{a}_{\scriptscriptstyle{0}} - \hat{a}^{\dag}_{\scriptscriptstyle{0}}\hat{a}_{\scriptscriptstyle{-1}} + \hat{a}^{\dag}_{\scriptscriptstyle{0}}\hat{a}_{\scriptscriptstyle{+1}} - \hat{a}^{\dag}_{\scriptscriptstyle{+1}}\hat{a}_{\scriptscriptstyle{0}}\right), \\ 
	\nonumber
	\hat{J}_{z} &\ =\  \hat{a}^{\dag}_{\scriptscriptstyle{+1}}\hat{a}_{\scriptscriptstyle{+1}} - \hat{a}^{\dag}_{\scriptscriptstyle{-1}}\hat{a}_{\scriptscriptstyle{-1}},\\
\end{align}
with the total spin vector operator $\hat{J}^2=\hat{J}_x^2+\hat{J}_y^2+\hat{J}_z^2$. Moreover, the number operator in the $m_f$ Zeeman state is defined as $\hat{N}_{mf}=\hat{\Psi}_{mf}^{\dagger}\hat{\Psi}_{mf}$ equivalent to $\hat{a}_{mf}^{\dagger}\hat{a}_{mf}$.

\section{Numerical method}\label{app2}

In order to diagonalize the Hamiltonian (\ref{H}) one can use either the Fock or the Dicke basis. In the following, we give the parametrization of the Hamiltonian (\ref{H}) for both of these basis. 

\subsection{Fock basis}
For the diagonalization of (\ref{H}), it is convenient to use the Fock basis, which is equivalent to mean-field ground state basis used in Ref.~\cite{Zhang2003}. In this case, the occupation number of particles in each Zeeman mode sub-level is used as Hamiltonian basis. We have used the Fock basis based with the following parametrization
\begin{eqnarray}
|k\rangle=|N_{+1},N_0,N_{-1}\rangle=|k,N+M-2k,k-M\rangle ,
\end{eqnarray}
which leads to the bounds on $k$ as
\begin{eqnarray}
k_{\rm min}&=&{\rm max}[0,M/2,M],\\
k_{\rm max}&=&{\rm min}[N,(M+N)/2,N+M].
\end{eqnarray}
Subsequently, we builds up the Hamiltonian in this basis and numerically diagonalize it to obtain the ground state. The resulting Hamiltonian has a block-diagonal structure in the basis with the size $dim=k_{max}-k_{min}+1$. In the extreme limits of $M=0$ and $M=N$ the size of the block is $dim=N/2-M+1$ and $1$, respectively, i. e. the size of blocks decreases for larger magnetization values.

\subsection{Dicke basis}

In order to use the perturbation theory (Sec. \ref{sec2} and Appendix \ref{app2}) , it is more straightforward to diagonalize the Hamiltonian in the Dicke basis~\cite{Gerbier2012}. To this end, let us  suppose first that there is no external transverse magnetic field ($q=0$) and then the Hamiltonian (\ref{H}) reduces to the form of $\frac{c}{2N}\hat{J}^2$. The respective eigenstates are $|N,\mathcal{J},M\rangle$ and their corresponding eigenvalues $\frac{c}{2N}\mathcal{J}(\mathcal{J}+1)$, where $\mathcal{J}$ and $M$ represent the total spin number and magnetization ($\hat{J}_z$ eigenvalues), respectively. Each state has $2\mathcal{J}+1$ degeneracy. Now, if $q\neq{0}$ due to $[\hat{J}_z,\hat{N}_0]=0$ the magnetization is still a good quantum number and therefore one can diagonalize $H$ in each block of fixed magnetization $M$. The Dicke basis may be defined in terms of Fock basis as \cite{Sarlo2013}
\begin{eqnarray}
|N,\mathcal{J},M\rangle=\frac{1}{\mathcal{N}}(\hat{J}^{(-)})^P(\hat{A}^\dagger)^Q(\hat{a}_{+1})^\mathcal{J}|\text{vac}\rangle ,
\end{eqnarray}
where $P=\mathcal{J}-M$, $2Q=N-\mathcal{J}$, $\hat{J}^{(-)}=\sqrt{2}(\hat{a}^\dagger_{-1}\hat{a}_0+\hat{a}^\dagger_{0}\hat{a}_1)$ is spin lowering operator and $\hat{A}^\dagger=(\hat{a}_0^\dagger)^2-2\hat{a}_{-1}^\dagger \hat{a}_{+1}^\dagger$ is the singlet spin operator with the following normalization factor
\begin{eqnarray}
\mathcal{N}=\frac{\mathcal{J}!(N-\mathcal{J})(N+\mathcal{J}+1)!!(\mathcal{J}-M)!(2\mathcal{J})!}{(2\mathcal{J}+1)!!(\mathcal{J}+M)!}
\end{eqnarray}
with $!!$ being the double fractional. By acting $\hat{a}_0$ on the Dicke states we have
\begin{eqnarray}\nonumber
\hat{a}_0|N,\mathcal{J},M\rangle&=&\sqrt{A_-(N,\mathcal{J},M)}|N-1,\mathcal{J}-1,M\rangle\\ \nonumber
&+&\sqrt{A_+(N,\mathcal{J},M)}|N+1,\mathcal{J}+1,M\rangle\\
\end{eqnarray}
with
\begin{eqnarray}\nonumber
A_-(N,\mathcal{J},M)&=&\frac{(\mathcal{J}^2-M^2)(N+\mathcal{J}+1)}{(2\mathcal{J}-1)(2\mathcal{J}+1)},\\
A_+(N,\mathcal{J},M)&=&\frac{((\mathcal{J}+1)^2-M^2)(N-\mathcal{J})}{(2\mathcal{J}+1)(2\mathcal{J}+3)}.
\end{eqnarray}
Note, that due to $2Q=N-\mathcal{J}$, the eigenstates might have the even or odd parities depending on the number of particles.

The state of the system can be considered in the Dicke basis as 
\begin{equation}
|\psi\rangle=\sum_{\mathcal{J}=|M|}^N C_{\mathcal{J},M}|N,\mathcal{J},M\rangle
\end{equation}
In order to build the time-independent Schr\"odinger equation of (\ref{H}),  $\hat{H}|\psi\rangle=E|\psi\rangle$, one needs 
the matrix elements such as $\langle N, \mathcal{J}, M|\hat{N}_0|N,\mathcal{J}',M\rangle$. It has been proved that the only non-zero elements are with $\mathcal{J}'=\mathcal{J},\mathcal{J}\pm2$, and hence the Schr\"odinger equation leads to the following tridiagonal matrix form  \cite{Sarlo2013}
\begin{eqnarray}\nonumber
h^M_{\mathcal{J},\mathcal{J}+2}C_{\mathcal{J}+2,M}
+h^M_{\mathcal{J},\mathcal{J}-2}C_{\mathcal{J}-2,M}+h^M_{\mathcal{J},\mathcal{J}}C_{\mathcal{J},M}={E}C_{\mathcal{J},M}.\\
\end{eqnarray}
Here, ${E}$ refers to the eigenenergies and the respective coefficients are given as 
\begin{eqnarray}\nonumber
h_{\mathcal{J},\mathcal{J}}^{M}&=&\frac{c}{2N}\mathcal{J}(\mathcal{J}+1)-q\langle \mathcal{J}|\hat{N}_0|\mathcal{J}\rangle,\\ \nonumber
h_{\mathcal{J},\mathcal{J}+2}^{M}&=&-q\langle \mathcal{J}+2|\hat{N}_0|\mathcal{J}\rangle,\\
h_{\mathcal{J},\mathcal{J}-2}^M&=&-q\langle \mathcal{J}-2|\hat{N}_0|\mathcal{J}\rangle ,
\end{eqnarray}
and
\begin{eqnarray}\nonumber
\langle \mathcal{J}|\hat{N}_0|\mathcal{J}\rangle&=&A_+(N,\mathcal{J},M)+A_-(N,\mathcal{J},M)\\ \nonumber
\langle \mathcal{J}+2|\hat{N}_0|\mathcal{J}\rangle&=&\sqrt{A_{-}(N,\mathcal{J}+2,M)A_+(N,\mathcal{J},M)}\\ \nonumber
\langle \mathcal{J}-2|\hat{N}_0|\mathcal{J}\rangle &=&\sqrt{A_+(N,\mathcal{J}-2,M)A_-(N,\mathcal{J},M)},
\label{N0}
\end{eqnarray}
where we introduced notation $|\mathcal{J}\rangle= |N, \mathcal{J}, M\rangle$.

In the paper we have also used the following expressions involving $\hat{N}_0^2$
\begin{eqnarray}\nonumber
\langle \mathcal{J}|\hat{N}_0^2|\mathcal{J}\rangle &=& \nonumber [A_+(N,\mathcal{J},M)+A_-(N,\mathcal{J},M)]^2\\ \nonumber
&+&A_-(\mathcal{J}+2)A_+(\mathcal{J})
+A_-(\mathcal{J})A_+(\mathcal{J}-2), \\ \nonumber
\langle \mathcal{J}|\hat{N}_0^2|\mathcal{J}+2\rangle &=& \nonumber
\sqrt{(A_+(\mathcal{J})A_-(\mathcal{J}+2)}
+A_+(\mathcal{J})+A_-(\mathcal{J})\\ \nonumber
&+&A_+(\mathcal{J}+2)+A_-(\mathcal{J}+2)),\\
\langle \mathcal{J}|\hat{N}_0^2|\mathcal{J}-2\rangle &=& \nonumber
\sqrt{(A_-(\mathcal{J})A_+(\mathcal{J}-2)}+A_+(\mathcal{J})+A_-(\mathcal{J})\\ \nonumber
&+&A_+(\mathcal{J}-2)+A_-(\mathcal{J}-2)),\\ \nonumber
\langle \mathcal{J}|\hat{N}_0^2|\mathcal{J}+4\rangle &=& \nonumber
\sqrt{A_+(\mathcal{J})A_-(\mathcal{J}+2)A_+(\mathcal{J}+2)A_-(\mathcal{J}+4)},\\ \nonumber
\langle \mathcal{J}|\hat{N}_0^2|\mathcal{J}-4\rangle &=& \nonumber
\sqrt{A_-(\mathcal{J})A_+(\mathcal{J}-2)A_-(\mathcal{J}-2)A_+(\mathcal{J}-4)}.\\ 
\label{N02}
\end{eqnarray}
The fact that $\hat{N}_0$ only connects eigenstates with $\mathcal{J}$ and $\mathcal{J}\pm{2}$ results in the Hamiltonian eigenstates having even or odd parity~\cite{Sarlo2013,Niezgoda2019}.

\section{Eigenstates and eigenvalues of the Hamiltonian (\ref{H}): Perturbation theory}\label{app3}

In order to obtain the analytical expressions for the QFI for zero magnetization, we use the second-order perturbation theory to find eigenstates and eigenenergies of Hamiltonian (\ref{H}). To this end, as mentioned in Sec. \ref{sec3}, we employ the second-order perturbation theory in the Dicke basis for small values of $q$. Let us take the unperturbed Hamiltonian $\hat{H}_0$ which satisfies $\hat{H}_0|\psi_n^{(0)}\rangle=E_n^{(0)}|\psi_n^{(0)}\rangle$
where $E_n^{(0)}$ and $|\psi_{n}(0)\rangle$ are the
eigenvalues and eigenstates, respectively.
Based on the second-order perturbation theory, the eigenbasis of perturbed Schr\"odinger equation $(\hat{H}_0+q\hat{H}_q)|\tilde{\psi}_n\rangle=\tilde{E}_n|\tilde{\psi}_n\rangle$ can be calculated as~\cite{Sakurai1994}
\begin{eqnarray}\nonumber
\tilde{E}_n&=&{E}_n^{(0)}+q{E}_n^{(1)}+q^2{E}_n^{(2)}+\mathcal{O}(q^3),\\ \nonumber
|\tilde{\psi}_n\rangle&=&|\psi_n^{(0)}\rangle+q|\psi_n^{(1)}\rangle+q^2|\psi_n^{(2)}\rangle+\mathcal{O}(q^3),
\end{eqnarray}

\noindent where we can find the corrections to the eigenenergies via 
\begin{eqnarray}\nonumber
\tilde{E}_n^{(1)}&=-&\langle\psi_n^{(0)}|\hat{N}_0|{\psi}_n^{(0)}\rangle,\\ \label{EV}
\tilde{E}_n^{(2)}&=&\sum_{m\neq n}\frac{|\langle \psi_m^{(0)}|\hat{N}_0|\psi_n^{(0)}\rangle|^2}{E_n^{(0)}-E_{m}^{(0)}},
\end{eqnarray}

\noindent and to the eigenstates as

\begin{eqnarray}\nonumber
|{\psi}_n^{(1)}\rangle&=&\sum_{m\neq n}\frac{\langle \psi_m^{(0)}|\hat{N}_0|\psi_n^{(0)}\rangle}{E_n^{(0)}-E_{m}^{(0)}}|\psi_n^{(0)}\rangle,\\ \nonumber
|{\psi}_n^{(2)}\rangle&=&\sum_{k\neq n}\sum_{m\neq n}\frac{\langle \psi_m^{(0)}|\hat{N}_0|\psi_k^{(0)}\rangle\langle \psi_k^{(0)}|\hat{N}_0|\psi_n^{(0)}\rangle}{(E_n^{(0)}-E_{l}^{(0)})(E_n^{(0)}-E_{m}^{(0)})}|\psi_n^{(0)}\rangle \\ \nonumber
&-&\sum_{k\neq n}
\frac{\langle \psi_n^{(0)}|\hat{N}_0|\psi_n^{(0)}\rangle\langle \psi_k^{(0)}|\hat{N}_0|\psi_n^{(0)}\rangle}{(E_n^{(0)}-E_{k}^{(0)})^2}|\psi_k^{(0)}\rangle\\ \nonumber
&-&\frac{1}{2}\sum_{k\neq n}
\frac{\langle \psi_n^{(0)}|\hat{N}_0|\psi_k^{(0)}\rangle\langle \psi_k^{(0)}|\hat{N}_0|\psi_n^{(0)}\rangle}{(E_n^{(0)}-E_{k}^{(0)})^2}|\psi_n^{(0)}\rangle.\\ \label{ES}
\end{eqnarray}

\noindent Now, lets consider the case of $\hat{H}_0=\frac{c}{2N}\hat{J}^2$, the eigenstates being given by the  conventional Dicke states $|N,\mathcal{J},M\rangle$ with  eigenvalues $\frac{c}{2N}\mathcal{J}(\mathcal{J}+1)$. Particularly, the ground state in this case is known as $|N,\mathcal{J}=0,M=0\rangle$ and the first excited state as $|N,\mathcal{J}=2,M=0\rangle$ (due to parity condition). Using Eqs.
(\ref{EV}) and (\ref{ES}) we can extract the perturbative results for the ground and the first excited states of the Hamiltonian (\ref{H}), which read


\begin{align}
    |\tilde{\psi}_0\rangle&= 
    \left(1-q^2 \frac{2N^3(N+3)}{405} \right) |\mathcal{J}=0 \rangle \label{p_gs}  \\
    +& q\frac{2N\sqrt{N(N+3)}}{9\sqrt{5}}\left( 1 + q \frac{2N(2N+3)}{63} \right) |\mathcal{J}=2 \rangle \nonumber \\
    +& q^2 \frac{4N^2\sqrt{N(N+5)(N+3)(N-2)}}{1575} |\mathcal{J}=4 \rangle, \label{P_S}
\end{align}

\begin{align}
    &|\tilde{\psi}_2\rangle = 
    -q\frac{2N\sqrt{N(N+3)}}{9\sqrt{5}}\left( 1 + q \frac{2N(2N+3)}{63} \right) |\mathcal{J}=0 \rangle \nonumber \\
    &+ \left(1-q^2 \frac{2N^3(N+3)}{405} -q^2 \frac{8N^2(N+5)(N-2)}{12005}   \right) |\mathcal{J}=2 \rangle \nonumber \\
    &+q \frac{4N\sqrt{(N+5)(N-2)}}{49\sqrt{5}}\left( 1-q\frac{2N(2N+3)}{1617}\right)  |\mathcal{J}=4 \rangle \nonumber \\
    & + q^2 \frac{4\sqrt{5}N^2\sqrt{(N+7)(N+5)(N-2)(N-4)}}{4851\sqrt{13}}| \mathcal{J}=6 \rangle 
\end{align}

respectively, corresponding to the following eigenenergies

\begin{eqnarray}\nonumber
\tilde{E}_0&=& -q\frac{N}{3}- q^2\frac{4N^2(N+3) }{135},\\
\nonumber
\tilde{E}_2&=&\frac{3}{N} - q\frac{11N + 6}{21}\\ \nonumber 
& + &q^2\left[ \frac{4N^2(N+3)}{135} - \frac{16N(N+5)(N-2)}{1715} \right]. \label{P_V}
\end{eqnarray}

\noindent Note, the energy gap scales as $3/N$ and it is in agreement with the exact numerical results~\cite{Gerbier2012}. In order to find the QFI value, it suffices to consider only the two lowest energy states of the Hamiltonian (\ref{H}).
A similar dependence of QFI on the value of two lowest lying energy states has been observed in the Lipkin-Meshkov-Glick model~\cite{Paris2014}. Consequently, we derive an analytical formula for the QFI, making use of (\ref{P_S}) and (\ref{P_V}). The final expression gets unwieldy and we have not the brought explicit forms here. Instead in Fig.~\ref{fig:fig3} we show the QFI for $N=1000$ and $M=0$ using both the exact numerical and perturbative results which demonstrate good agreement for $q\sim 0$. 
In particular, it is interesting to note that the maximal value of the QFI around criticality can be derived easily by inserting (\ref{p_gs}) into (\ref{QFI1}) which gives $\frac{16}{405}N^4\approx 0.04N^4$. The exponent is in a very good agreement with the numerical results obtained by the exact diagonalization method given in section \ref{sec3a}. \footnote{Alternatively, one can make use of (\ref{P_S}) and (\ref{P_V}) up to the second order of $q$ into (\ref{QFI}) and then let $q\rightarrow 0$. This yields the scaling of $\frac{16}{405}N^4\approx 0.04N^4$ for quantum Fisher information and around critical point.}  The scaling of the precision versus $N$ might be evaluated based on the error-propagation formula (\ref{delta}). Using (\ref{p_gs}), (\ref{N0}) and (\ref{N02})
for the first and second moments of $\hat{N}_0$
around QCP gives

\begin{eqnarray}
\Delta^2 \hat{N}_0 &=& \frac{4N(N+3)}{45}\\
\partial_q\ave{\hat{N}_0}&=&\frac{8}{135}{N^2(N+3)}.
\end{eqnarray}

\noindent The value of the variance which confirms the result of Ref. \cite{Gerbier2012}
refers to the super-Poissonian statistics of the BEC in the single state. Using the precision, leads to  $\frac{16N^3(N+3)}{405}=0.04N^4$ which are in an excellent agreement with the numerical results presented in Sec. \ref{sec3}.

\begin{figure}[]
\centering
\begin{picture}(0,110)
\put(-125,0)
{\includegraphics[width=0.5\linewidth]{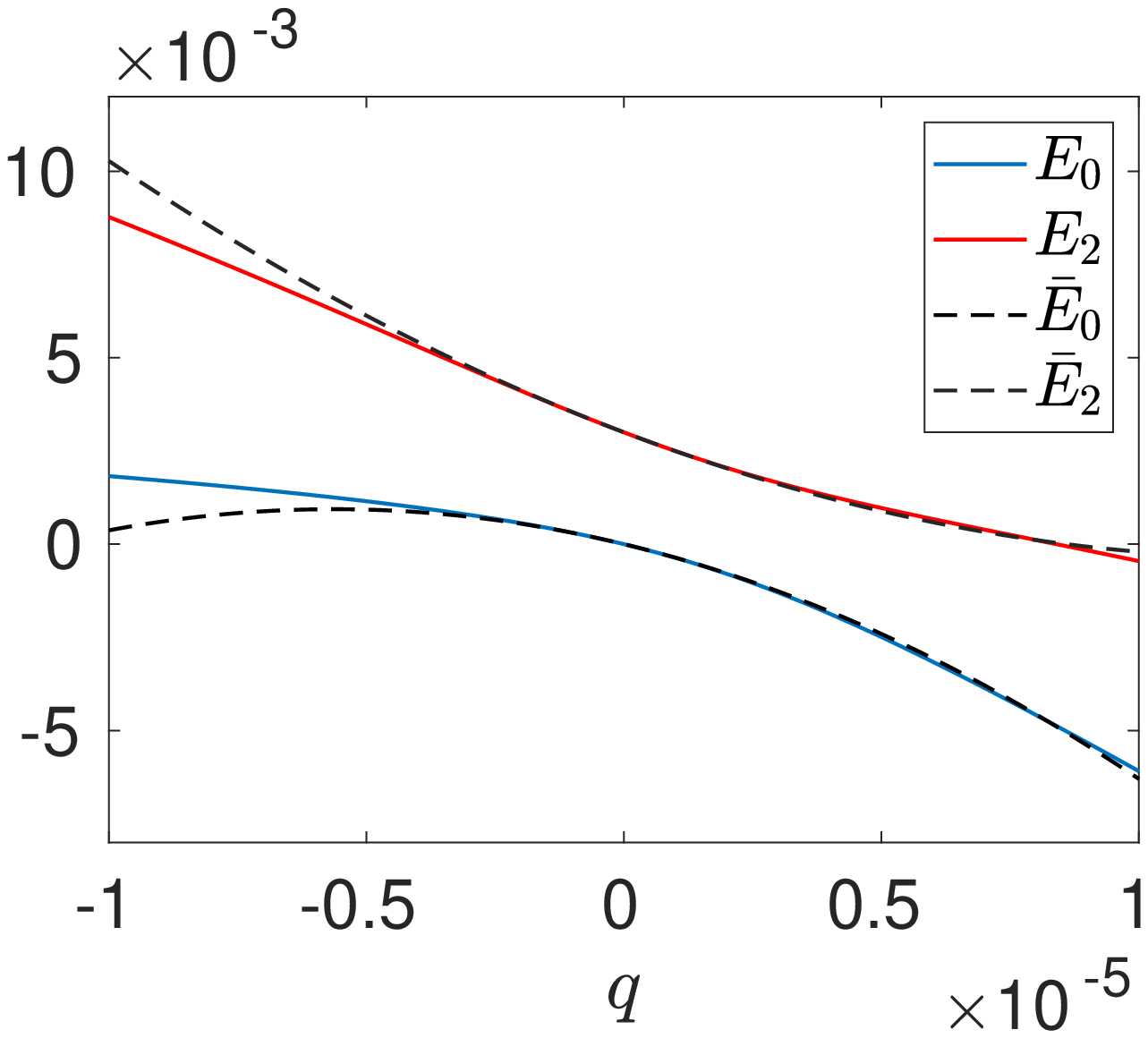}}
\put(-120,98){(a)}
\put(5,0)
{\includegraphics[width=0.5\linewidth]{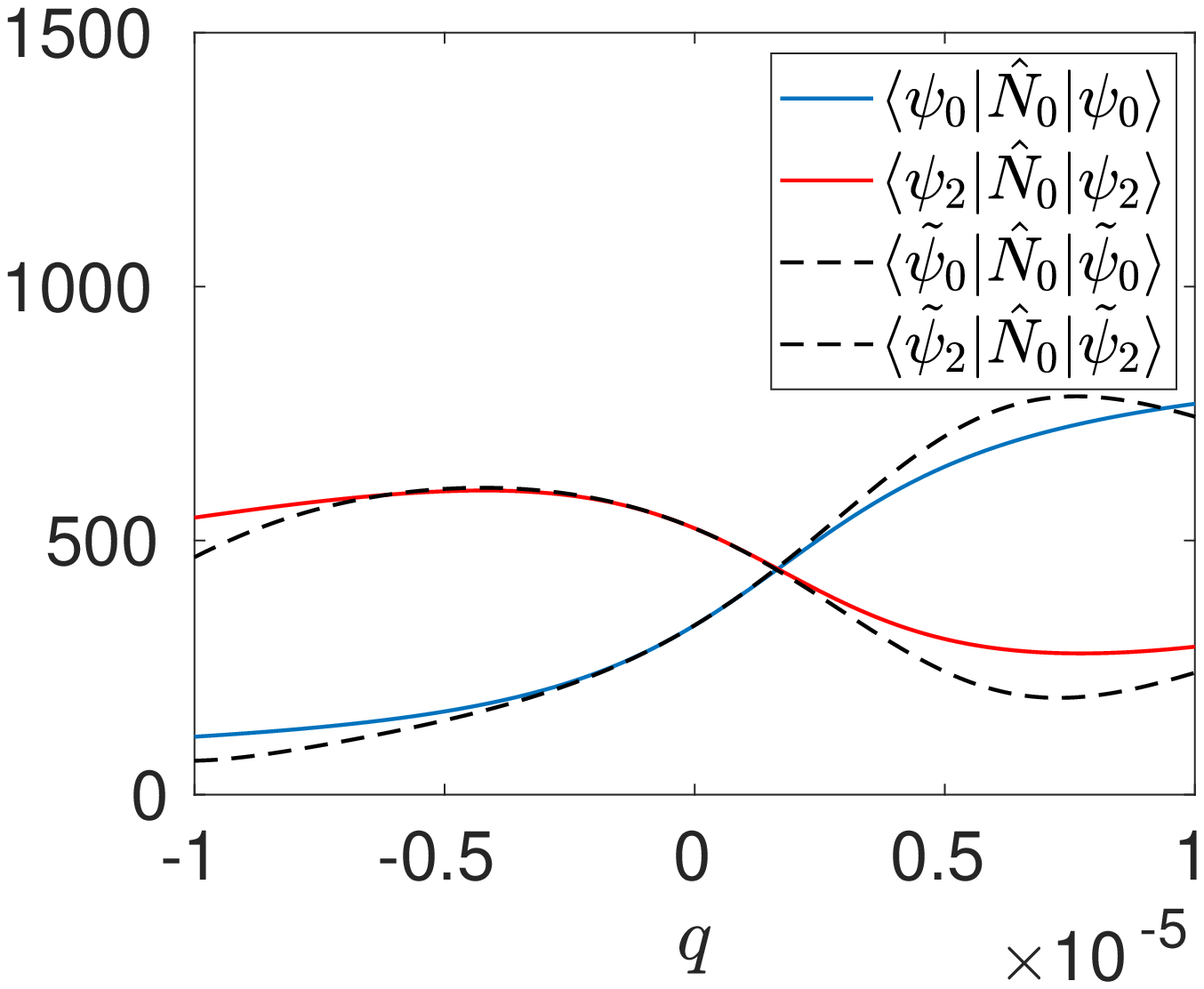}}
\put(25,98){(b)}
\end{picture}
\caption{(a) The energy of the ground $E_0$ (the blue solid line) and the first excited $E_2$ (the red solid line) states of the Hamiltonian (\ref{H}) from exact diagonalization method. The approximated results from perturbation theory  are marked by the corresponding dashed black lines. Here, the total atom number is  $N=1000$ and $M=0$. (b) The average value of population in the $m_f=0$ Zeeman level using the ground state $\langle\psi_0|\hat{N}_0|\psi_0\rangle$ (the blue solid line) and the first excited state $\langle\psi_2|\hat{N}_0|\psi_2\rangle$ (the red solid line). The corresponding approximate values are given with the dashed black lines as $\langle\tilde{\psi}_0|\hat{N}_0|\tilde{\psi}_0\rangle$ and $\langle\tilde{\psi}_2|\hat{N}_0|\tilde{\psi}_2\rangle$.}
\label{fig:fig12}
\end{figure}

\noindent In Fig. \ref{fig:fig12}(a) we have shown the eigenenergies of (\ref{P_V}) versus $q$ using both the exact numerical and the approximate approaches. Moreover in order to check the validity of our perturbation approach in Fig. \ref{fig:fig12}(b) we have presented the average value of $\hat{N}_0$ over both the ground and the first excited states, using both perturbative and exact numerical diagonalization of Hamiltonian (\ref{H}). There are good agreements in limits of validity. As we see, for $q=0$, the results give the singlet state which is specified by $\ave{\hat{N}_0}=N/3$ \cite{Bigelow1998}.

\bibliography{phase_bib}

\end{document}